\documentclass[aps,prd,showpacs,twocolumn,10pt,superscriptaddress,floatfix,nofootinbib,longbibliography]{revtex4-1}

\usepackage{amssymb,amsfonts,amsmath,bm,bbm}
\usepackage{graphicx}
\usepackage{dcolumn}
\usepackage[
colorlinks=true,
linkcolor=blue,  
breaklinks=true,
urlcolor=blue,
citecolor=blue]{hyperref}
\usepackage{bbold}
\usepackage{epstopdf}

\newcommand{\HNUST}{\affiliation{
Hunan Provincial Key Laboratory of Intelligent Sensors and Advanced Sensor Materials, \\ School of Physics and Electronics, Hunan University of Science and Technology, Xiangtan 411201, China}}

\newcommand{\BU}{\affiliation{
School of Physics, Beihang University, Beijing 102206, China}}

\newcommand{\USC}{\affiliation{
School of Nuclear Science and Technology, University of South China, Hengyang 421001, China}}

\newcommand{\FDU}{\affiliation{
Physics Department and Center for Particle Physics and Field Theory, Fudan University, Shanghai 200438, China}}

\newcommand{\HU}{\affiliation{
School of Physics and Electronics, Hunan University, Changsha 410082, China}}

\begin{document}

\title{
QCD topology and axion properties in an isotropic hot and dense medium
}

\author{Hong-Fang Gong}\thanks{These authors contributed equally to this work.}
\HNUST

\author{Qi Lu}\thanks{These authors contributed equally to this work.}
\BU\HNUST

\author{Zhen-Yan Lu}
\email[Corresponding author:~]{luzhenyan@hnust.edu.cn}\HNUST

\author{Lu-Meng Liu}\email{liulumeng@fudan.edu.cn}\FDU

\author{Xun Chen}\email{chenxun@usc.edu.cn}\USC

\author{Shu-Peng Wang}\HNUST\HU

\date{\today}

\begin{abstract}
We study the QCD topology and axion properties at finite temperature and chemical potential in the framework of the two-flavor Nambu$-$Jona-Lasinio model. We find that the behaviors of the two lowest cumulants of the QCD topological charge distribution and axion properties are highly sensitive to the critical behavior of the chiral phase transition. In particular, the topological susceptibility and the axion mass follow the response of the chiral condensate to temperature and chemical potential, showing that both quantities decrease monotonically with the increment of temperature and/or chemical potential. However, it is important to note that the normalized fourth cumulant behaves differently depending on the temperature. At low temperatures, it is a non-monotonic function of the chemical potential, while at high temperatures, it monotonically decreases. Additionally, its value invariably approaches the asymptotic value of $b_2^{\text {inst }}=-1/12$, predicted by the dilute instanton gas model.
We also observe that with the increase in chemical potential at relatively low temperatures, the axion self-coupling constant exhibits a sharp peak around the critical point, which can even be more than twice its vacuum value. After that, the self-coupling drops sharply to a much lower value than its vacuum value, eventually approaching zero in the high chemical potential limit. The finding that the axion self-coupling constant is significantly enhanced in high-density environments near the chiral phase transition could lead to the creation or enhancement of an axion Bose-Einstein condensate in compact astrophysical objects.
\end{abstract}

\maketitle

\section{Introduction} \label{sec:INTRODUCTION}

Within the domain
of quantum physics, quantum chromodynamics (QCD) serves as a pivotal
theory delineating the intricate dynamics of the strong force~\cite{Gross-2022hyw}. This non-abelian gauge theory, expressed through the Lagrangian formalism, highlights quarks as the essential elements.
Remarkably, the presence of instantons~\cite{81Gross.Pisarski.ea43-43RMP,86Hooft357-387PR,98Schdotafer.Shuryak323-426RMP},
which serves as
a resolution to the U(1)$_A$ problem~\cite{76Hooft8-11PRL,76Hooft3432-3450PRD}, implies the necessity of incorporating a topological $\theta$-term into the QCD Lagrangian.
In this case, the QCD vacuum is nontrivial, which exhibits topological properties and is characterized by CP-even topological cumulants~\cite{Guo-2015oxa,Kawaguchi-2023olk}.
The $\theta$-term does not contribute to the dynamics of the theory,
however, it might give rise to the electric dipole moment of neutrons
~\cite{79Crewther.DiVecchia.ea123-123PLB}. 
Experimental measurements and lattice calculations
reveal that the value of $\theta$ is unnaturally small~\cite{Baker-2006ts,09Griffith.Swallows.ea101601-101601PRL,15Parker.others233002-233002PRL,16Graner.Chen.ea161601-161601PRL,Guo-2015tla}, $|\theta|\lesssim10^{-10}$, which results in the known puzzle called the strong CP problem.

Among the proposed strong CP problem solutions, the Peccei--Quinn (PQ) mechanism~\cite{77Peccei.Quinn1440-1443PRL,77Peccei.Quinn1791-1797PRD}, which introduces a global U(1) symmetry, called PQ symmetry, seems to be the most attractive one.
In this scenario, the axion appears as the corresponding Goldstone boson ~\cite{78Weinberg223-226PRL,78Wilczek279-282PRL} from the spontaneous breaking of the PQ symmetry.
Since the axion was proposed as a solution to the strong CP problem, there has been constant interest in studying its properties~\cite{Kim-2008hd,Caputo-2024oqc,13Kawasaki.Nakayama69-95ARNPS,Duffy-2009ig,16Marsh1-79PR,DiLuzio-2020wdo,DiLuzio-2021ysg,Choi-2020rgn} and searching for it~\cite{03Bradley.Clarke.ea777-817RMP,Jaeckel-2010ni,Ringwald-2012hr,15Graham.Irastorza.ea485-514ARNPS,Irastorza-2018dyq,Sikivie-2020zpn,Sikivie-2020zpn,Chen-2023jki,Semertzidis-2021rxs}.
Due to its light mass and weak interaction with the Standard Model particles~\cite{Antel-2023hkf}, the axion also constitutes an attractive candidate for dark matter~\cite{83Preskill.Wise.ea127-132PLB,Abbott-1982af,83DinePLB,OHare-2024nmr}.

With the present manuscript, we aim to
study how the cumulants of the QCD topological charge distribution and axion properties change with quark chemical potential and temperature,
which are of significant relevance for astrophysical systems, in particular
those with finite baryonic densities~\cite{Balkin-2020dsr,Anzuini-2023whm,Anzuini-2022bqd,Ferrer-2024xwu,Hook-2017psm,Lopes-2022efy,Yadav-2024xob,Cavan-Piton-2024ayu,Song-2024rru,Murgana-2024djt} and a hot medium~\cite{Salvio-2013iaa,DEramo-2021lgb,18Ferreira.Notari191301-191301PRL,Bandyopadhyay-2019pml,Notari-2022ffe,DEramo-2021psx,Ferreira-2020bpb,DEramo-2022nvb,DEramo-2018vss,Badziak-2024szg,Wang-2023xny}.
Nevertheless, when the quark chemical potential is finite, lattice QCD simulations with three colors encounter issues related to the sign problem~\cite{Splittorff-2007ck}. Consequently, the direct application of first-principle numerical computations in baryonic matter
confronts a great challenge.
For this reason, our understanding of QCD at low-temperature and
high-density regimes is still limited compared to that in hot medium.
Additionally,
we note that the QCD coupling is running,
and in the low energy regime, it becomes large enough to make it difficult to use perturbative QCD~\cite{Prosperi-2006hx,Deur-2016tte}.
As such, it becomes imperative for us to find ways to develop non-perturbative approaches, such as effective QCD models, to address issues in QCD.

The Nambu$-$Jona-Lasinio (NJL) model~\cite{Nambu-1961fr,61Nambu.Jona-Lasinio345-358PR,94Hatsuda.Kunihiro221-367PR,92Klevansky649-708RMP},
as one of the widely used effective QCD models,
provides a valuable tool for exploring the
properties of strongly interacting quark matter and
QCD phase transition~\cite{05Buballa-PhysRep,Volkov-2005kw}.
For instance, in our recent works, the
zero~\cite{Lu-2019diy} and finite temperature~\cite{Lu-2021hvw} results for the isospin imbalanced strongly interacting matter calculated in two-flavor NJL model are shown in good
agreement with the available lattice data as well as with the
results from chiral perturbation theory (CHPT).
In particular, the position of the peak found in the NJL model
for the ratio of energy density to the corresponding Stefan-Boltzmann limit~\cite{Lu-2019diy} was also found in good agreement with CHPT predication~\cite{16Carignano.Mammarella.ea51503-51503PRD} and lattice data~\cite{12Detmold.Orginos.ea54507-54507PRD}.
Furthermore, in the NJL model, the finding of the critical point for second order phase transition from the normal phase to the pion superfluid phase~\cite{05He.Jin.ea116001-116001PRD} at $\mu_I=m_\pi$ is quite consistent with the  lattice simulation~\cite{02Kogut.Sinclair34505-34505PRD,04Kogut.Sinclair94501-94501PRD} and the predictions from CHPT~\cite{01Son.Stephanov592-595PRL}.

Recently, the axion field was incorporated into the NJL model Lagrangian for the first time, and the effect of the chiral phase transition of QCD at zero chemical potential and finite temperature on the axion mass and self-coupling was calculated~\cite{Lu-2018ukl}. It was found that the axion mass decreases with temperature, following the topological susceptibility response, which agrees with previous results obtained within CHPT at low and intermediate temperatures. Especially the topological susceptibility calculated in the NJL model at zero temperature is in agreement with the results from CHPT~\cite{GrillidiCortona-2015jxo} and lattice data~\cite{16Borsanyi.others69-71N}.
As already explored in the previous works, the present model has provided successful descriptions for various QCD properties at finite temperatures and chemical potential.
Therefore,
we expect that the careful studies on the QCD topology and axion properties in a hot medium within the current theoretical framework will shed light on the understanding of the low energy properties of the QCD $\theta$-vacuum and axion in a finite baryonic density system at finite temperature.
We will see that the behaviors of the QCD topology and axion properties are significantly affected by the dense and hot QCD medium, especially near the critical point for the chiral phase transition.

The paper is organized as follows. In Sec.~\ref{sec:NJLMODEL}, we briefly review
the NJL model incorporating an axion field under finite temperature and chemical potential conditions.
In Sec.~\ref{sec:results}, we present our numerical results, in particular the results of temperature and finite chemical potential effects on the first two lowest cumulants of the QCD topological distribution and axion properties in an isotropic hot and dense medium. Finally, Sec.~\ref{sec:CONCLUSION} is devoted to the conclusions.

\section{NJL model} \label{sec:NJLMODEL}

The Lagrangian density of the two-flavor NJL model,
incorporating the U(1)$_A$ symmetry breaking term,
is given as
\begin{eqnarray}
\mathcal{L}=\bar{q}\left(i \gamma^\mu\partial_\mu+\mu \gamma_0-m_0\right) q
+\mathcal{L}_{\text {int }} ,
\end{eqnarray}
where $q\equiv(u,d)^T$ denotes the quark field matrix, $\mu$ the quark chemical potential, and $m_0$ is the degenerate current quark mass.
Additionally, the interaction term $\mathcal{L}_{\text {int}}$ is taken as
~\cite{09Boomsma.Boer34019-34019PRD,Das-2020pjg}
\begin{eqnarray}
\begin{aligned}
\mathcal{L}_{\text{int}} = &~ G_1\left[\left(\bar{q} \tau_a q\right)\left(\bar{q} \tau_a q\right)+\left(\bar{q}  i \tau_a\gamma_5 q\right)\left(\bar{q}  i \tau_a\gamma_5 q\right)\right] \\
& +8 G_2\left[e^{i \theta } \operatorname{det}\left(\bar{q}_R q_L\right)+e^{-i \theta } \operatorname{det}\left(\bar{q}_L q_R\right)\right],
\end{aligned}
\end{eqnarray}
which can be obtained by a chiral rotation of the quark fields in the path integral~\cite{76Hooft3432-3450PRD,86Hooft357-387PR,05Buballa-PhysRep}.
In the above equation, $\tau_a$ are matrices in the flavor space with $i=0,~1,~2,~3$; $\tau_0$ is the unit matrix and $\tau_i$ with $i=1,~2,~3$ are Pauli matrices. Here, the coupling constant $G_1$ serves as the governing factor for the $U(1)_A$ invariant interaction, while $G_2$ regulates the strength of the $U(1)_A$ breaking term.

Following the mean-field approximation, namely
\begin{eqnarray}
(\bar{q} q)^2 &\approx&  2(\bar{q} q)\langle \bar{q} q\rangle-\langle \bar{q} q\rangle^2,\\
(\bar{q}i\tau_a\gamma_5 q)^2 &\approx&
2(\bar{q}i\tau_a\gamma_5 q)\langle \bar{q} i\tau_a\gamma_5 q\rangle-\langle \bar{q} i\tau_a\gamma_5 q\rangle^2,
\end{eqnarray}
the thermodynamic potential at one loop can finally be given by~\cite{Lu-2018ukl}
\begin{eqnarray}\label{eq:Omegafull}
\Omega=\Omega_{\mathrm{mf}}+\Omega_q,
\end{eqnarray}
where the mean field contribution $\Omega_{\text{mf}}$ takes the form:
\begin{eqnarray}
\begin{aligned}
\Omega_{\mathrm{mf}}= & -G_2\left(\eta^2-\sigma^2\right) \cos \theta +G_1\left(\eta^2+\sigma^2\right) \\
& -2 G_2 \sigma \eta \sin \theta ,
\end{aligned}
\end{eqnarray}
with condensates $\sigma=\langle\bar{q} q\rangle$ and $\eta=\left\langle\bar{q} i \gamma_5 q\right\rangle$.
Furthermore, the quark loop contribution $\Omega_q$ is given by
\begin{eqnarray}\label{eq:OmegaQ}
\begin{aligned}
\Omega_q= & -2 N_c N_f T 
\int \frac{d^3 p}{(2 \pi)^3} 
\bigg\{\frac{E_p}{T} \\
&+
\ln 
\left[1+e^{-\left(E_p-\mu\right)/T}\right]
+\ln\left[1+e^{-\left(E_p+\mu\right)/T}\right]
\bigg\},
\end{aligned}
\end{eqnarray}
with the dispersion relations of quarks are given by
\begin{eqnarray}
E_p=\sqrt{p^2+\Delta^2}, \quad \Delta^2=\left(m_0+\alpha_0\right)^2+\beta_0^2,
\end{eqnarray}
with
\begin{eqnarray}
\begin{aligned}
& \alpha_0=-2\left(G_1+G_2 \cos \theta \right) \sigma+2 G_2 \eta \sin \theta , \\
& \beta_0=-2\left(G_1-G_2 \cos \theta \right) \eta+2 G_2 \sigma \sin \theta  .
\end{aligned}
\end{eqnarray}


The ground state of the
system at finite temperature and chemical potential is determined by minimizing the thermodynamic potential given in Eq.~(\ref{eq:Omegafull}) with respect to the condensates $\sigma$ and $\eta$, which leads to the following
gap equations
\begin{eqnarray}\label{eq:gapEquations2}
\frac{\partial \Omega}{\partial \sigma}\bigg|_{\sigma=\bar{\sigma}}=0, \quad \frac{\partial \Omega}{\partial \eta}\bigg|_{\eta=\bar{\eta}}=0,
\end{eqnarray}
with the solution $\sigma=\bar{\sigma}$, $\eta=\bar{\eta}$ corresponds to the global minimum of the thermodynamic potential of the system.
It is important to emphasize that for the phenomenological models, one should pay attention to addressing the thermodynamic inconsistency problem due to the consideration of the medium-dependent quark masses~\cite{Xu-2014zea,Ma-2023stj}. Fortunately, in the NJL model, thermodynamic consistency can be ensured
by the gap equations shown above.


\section{
Results and discussions} \label{sec:results}

Note that the first term
on the right-hand side of Eq.~(\ref{eq:OmegaQ}) exhibits ultraviolet divergence, which we manage by setting
a cutoff at $p=\Lambda$ for the integration. Thus, in the present calculation we have three parameters for the two-flavor NJL model: the current quark mass $m$, the coupling
constant $G$, and the cutoff $\Lambda$, which can be fixed by reproducing the empirical values of the pion mass $m_\pi=140.2$ MeV, the pion decay constant $f_\pi=92.6$ MeV, and the quark condensate in the vacuum $\sigma_0=2(-241.5~\text{MeV})^3$.
The obtained parameter values are~\cite{Lu-2018ukl} $\Lambda=590~\mathrm{MeV}$, $G_0=2.435/\Lambda^2$, $G_1=(1-c) G_0$, $G_2=c G_0$, $c=0.2$, $m_0=6~\mathrm{MeV}$.
Thereby,
for a given chemical potential and temperature, one can solve Eq.~(\ref{eq:gapEquations2}) numerically to obtain all the thermodynamic quantities of the system.

\subsection{Chiral condensate and QCD topology}

\begin{figure*}[htb]  
  \includegraphics[width=0.48\textwidth]{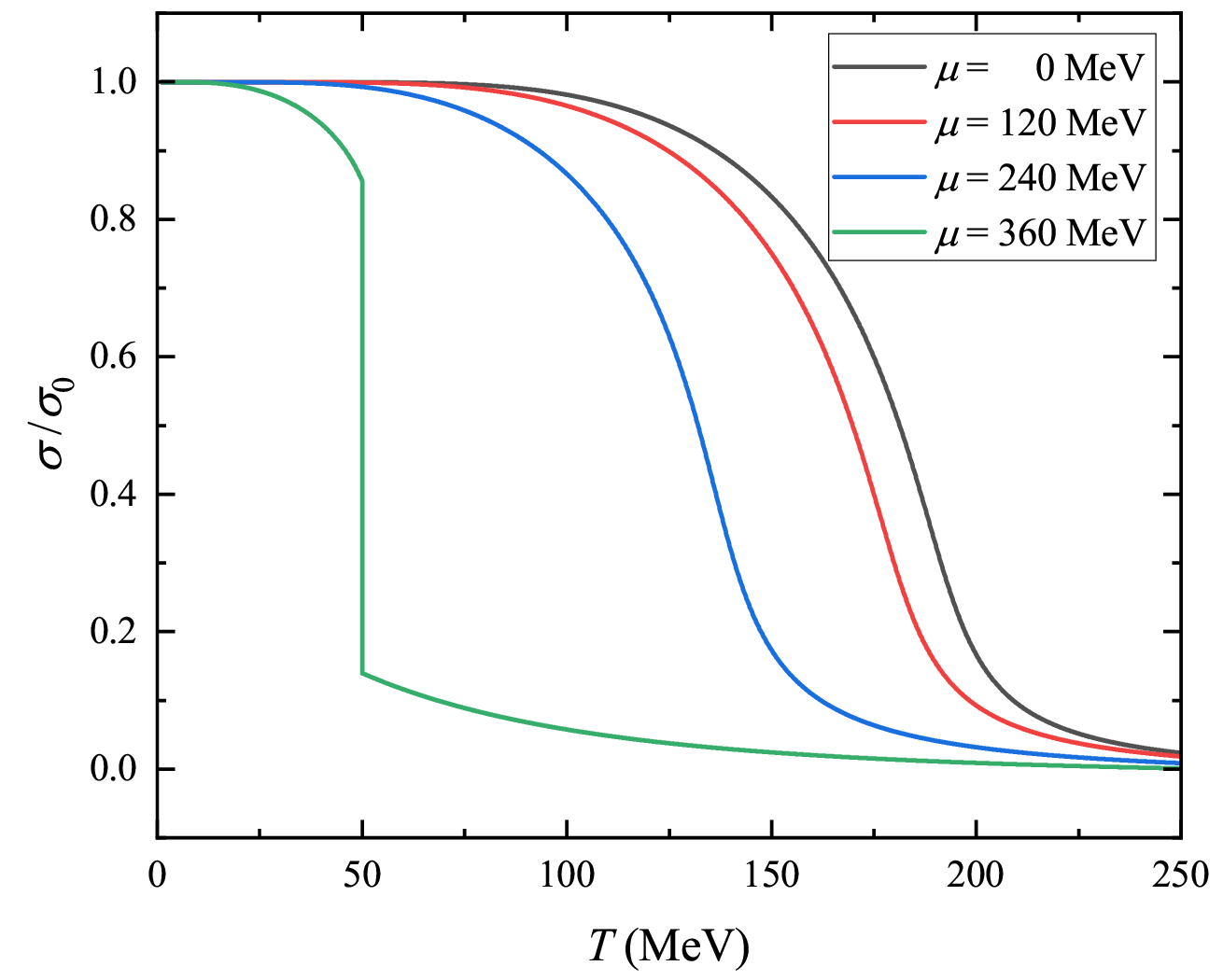}
   \includegraphics[width=0.48\textwidth]{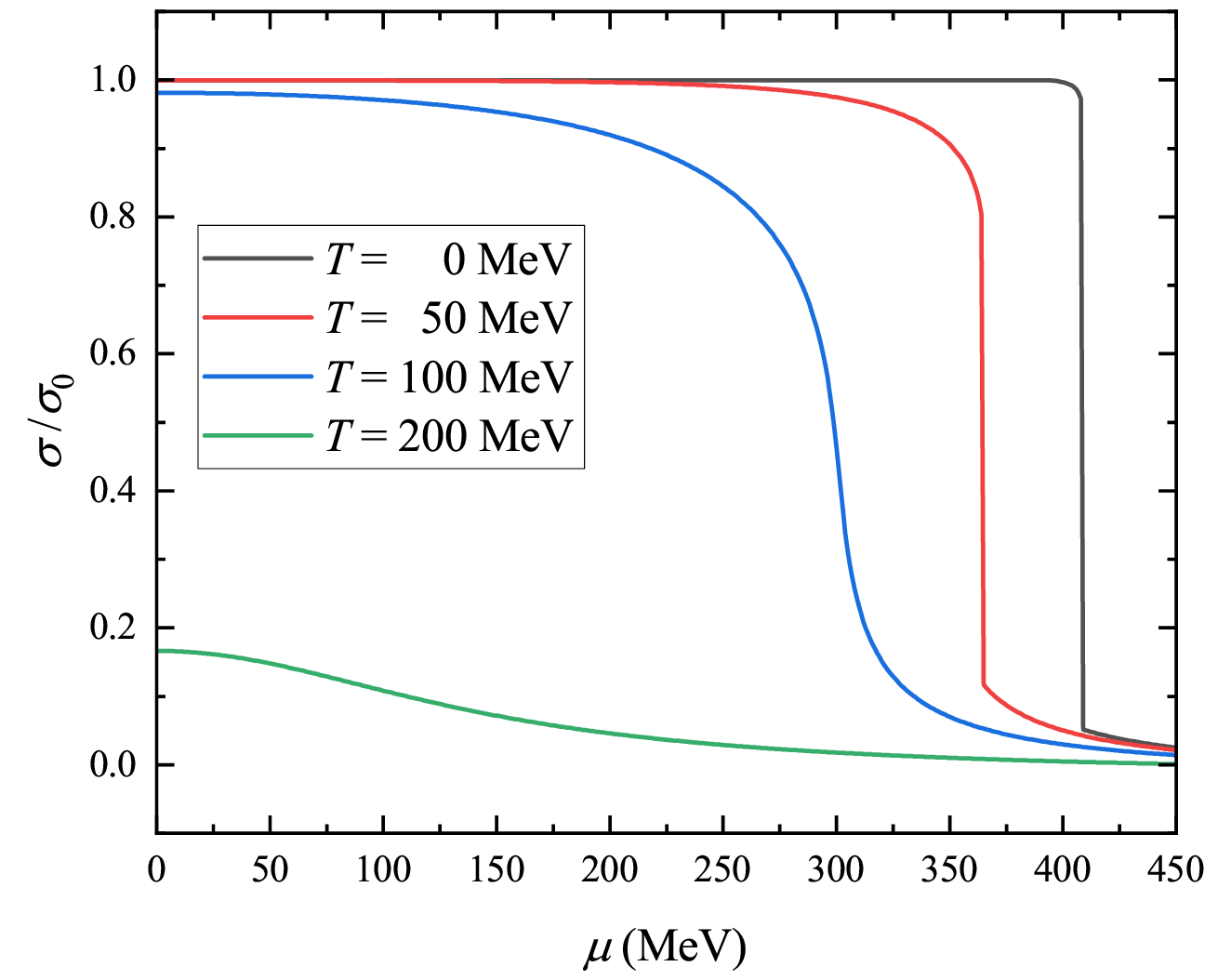}\\
  \caption{Variation of the chiral condensate, scaled by its value in the vacuum, with respect to the temperature at different chemical potentials  (left panel) and to the chemical potential at different temperatures  (right panel), respectively.
  }\label{fig:sigmaT}
\end{figure*}

The chiral condensate serves
as the order parameter to analyze the feature of the
chiral phase transition and can be numerically determined by the gap equations given in Eq.~(\ref{eq:gapEquations2}).
In the left panel of Fig.~\ref{fig:sigmaT}, we show the chiral condensate, scaled by its value in the vacuum, as a function of the temperature at different chemical potentials.
From right to left, the black, red, blue, and green curves correspond to the cases at $\mu=0,~120,~240,~360$ MeV, respectively. It can be seen that all the curves
remain constant at the vacuum value $\sigma_0$ until the chiral condensate starts to decrease at some value of the temperatures.
In particular, for the black, red, and blue solid curves with relatively low chemical potentials, it is evident that the chiral condensates decrease smoothly with the increment of the temperature, which indicates
that the restoration of the chiral symmetry for these three cases is always a chiral crossover.
However, with the green solid curve at $\mu = 360$ MeV, the chiral condensate shows a decrease at very low temperatures, followed by a sharp drop as the temperature rises to about $T\simeq 50$ MeV, and then a smooth decrease, indicating a first order phase transition occurring around the critical temperature. In other words, the chiral crossover has become a first order phase transition at sufficiently large chemical potentials~\cite{Karsch-2001cy,92Klevansky649-708RMP,Fodor-2001ye,Schaefer-2008hk}.

The right panel of Fig.~\ref{fig:sigmaT} shows how the scaled chiral condensate $\sigma/\sigma_0$
evolves with chemical potential. The curves in the figure correspond, from top to bottom, to the cases at $T=0,~50,~100,~200$ MeV, represented by black, red, blue, and green curves, respectively.
The green curve at the bottom starts to decrease from $0.17\sigma_0$ and shows an approximately linear decrease with
increasing chemical potential, while the other three curves
drop down in a very narrow range of chemical potentials, signaling the approximate restoration of chiral symmetry.
Although the top three curves with relatively low temperatures exhibit similar behavior at both low and high chemical potentials, they show different behavior at intermediate chemical potential values, particularly near the critical transition point.
More specifically, at zero temperature, the chiral condensate remains practically constant at the vacuum value $\sigma_0$ up to the threshold $\mu \simeq 409$ MeV. At this point, it drops to about $0.05\sigma_0$ and then continues to decrease smoothly as the chemical potential increases.
On the other hand, by increasing the temperature from $T=0$ MeV up to
$T = 50$ MeV,
the discontinuity point is
pushed towards smaller values of chemical potential, implying that the critical chemical potential decreases with the temperature.
One can also read from these two curves that in the case of $T=50$ MeV, the curve near the critical point becomes smoother than the one in the case of $T=0$: the temperature is to inhibit the
spontaneous breaking of chiral symmetry and reduces the chiral condensate of quarks.
In contrast to the other three curves, however, at $T = 100$ MeV the chiral condensate starts to decrease smoothly but not linearly with the increase of the chemical potential, which also indicates a chiral crossover, as does the green curve at $T=200$ MeV.


\begin{figure*}[htb]  
  \includegraphics[width=0.48\textwidth]{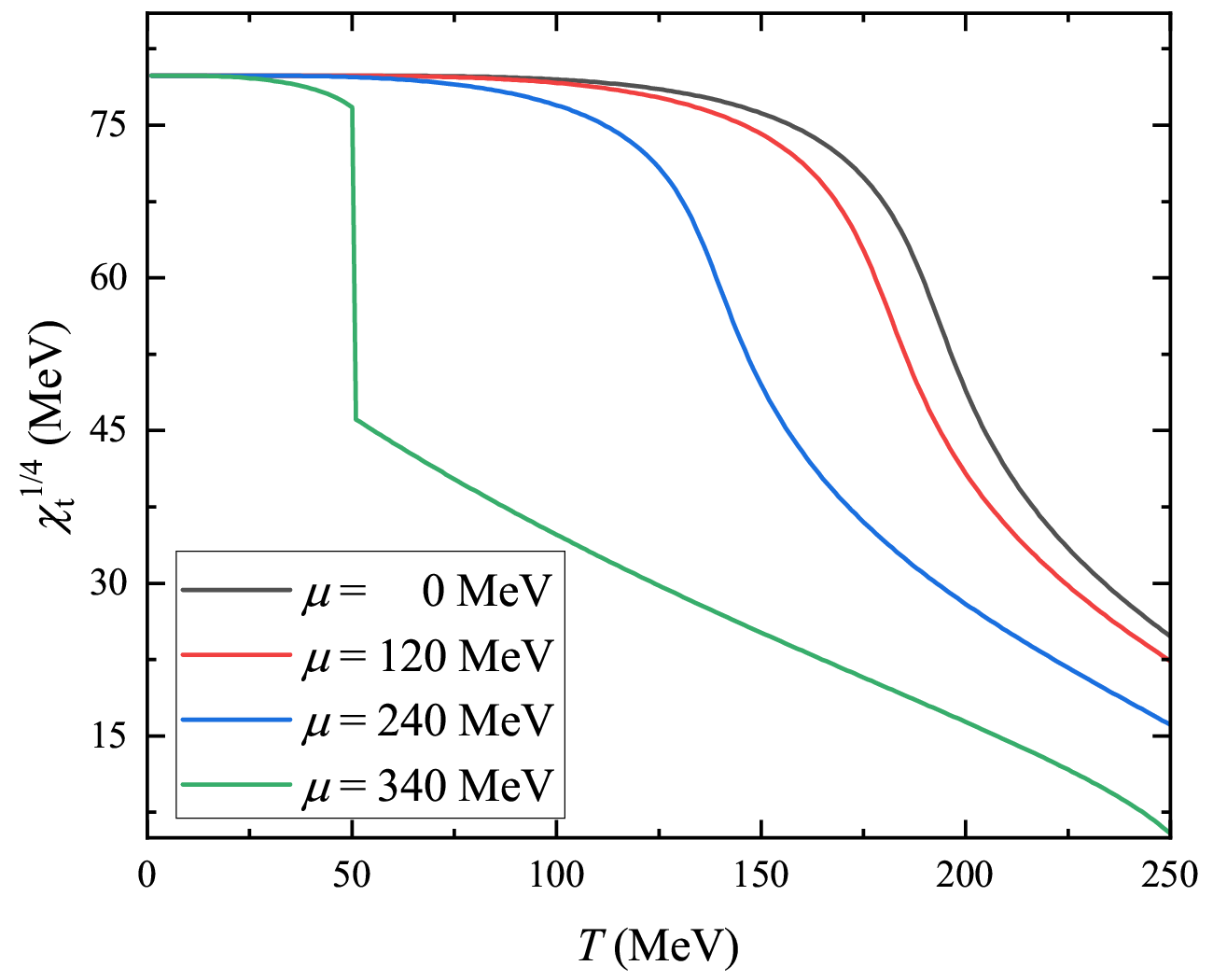}
   \includegraphics[width=0.48\textwidth]{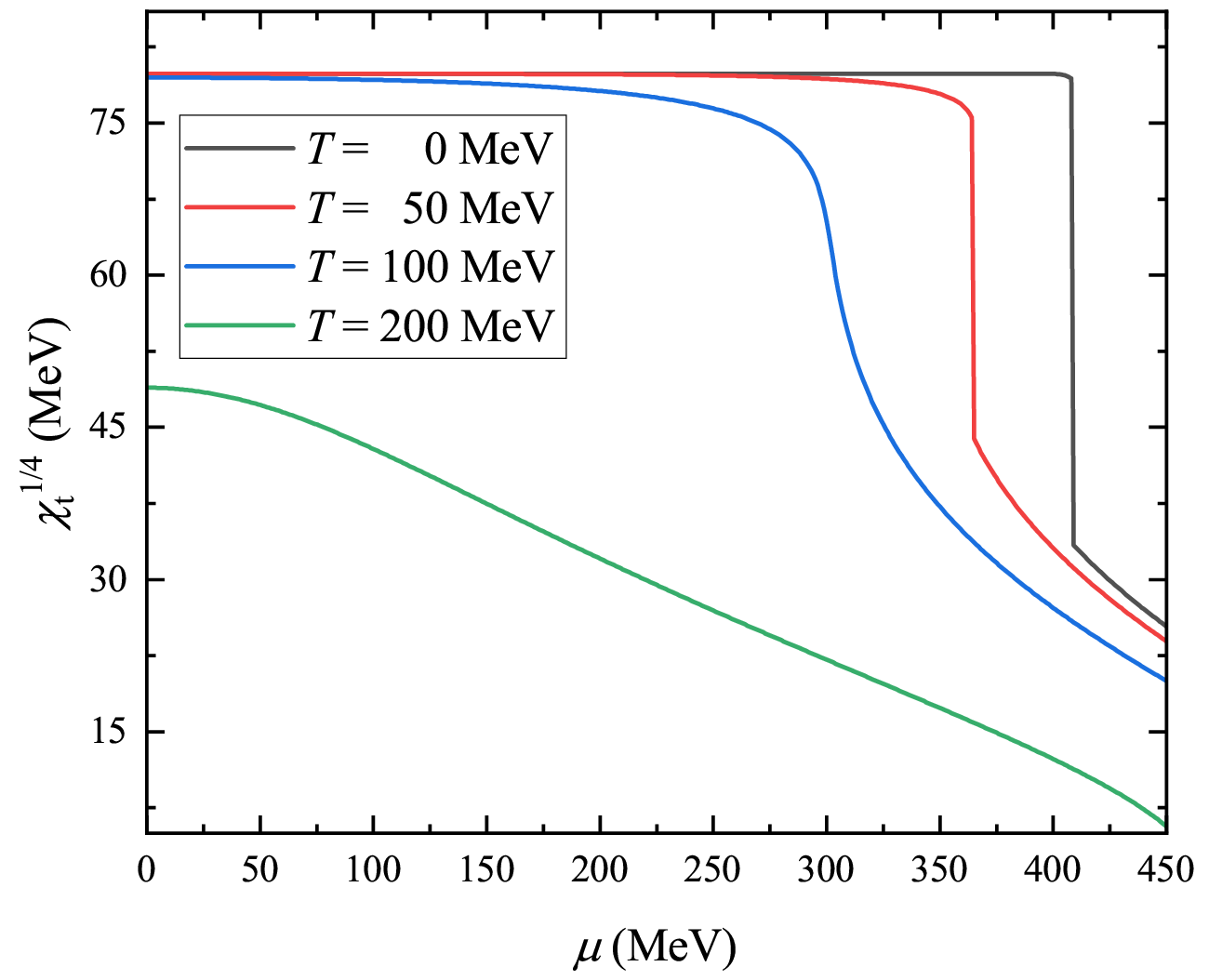}\\
  \caption{Variation of the topological susceptibility, scaled by its value in the vacuum, with respect to the temperature at different chemical potentials
  (left panel) and to the chemical potential at different temperatures  (right panel), respectively. Conventions for colors  are the same used in Fig.~\ref{fig:sigmaT}.
  }\label{fig:chiT}
\end{figure*}


The topological susceptibility is the leading cumulant of the QCD topological charge, and it has been precisely determined on the lattice and calculated from CHPT at zero temperature. Thus, the computations and applicability of the NJL model can, to some extent, be examined by comparison with those obtained from the lattice simulation and CHPT.
Now we turn to
the computation of the topological susceptibility $\chi_t$,
which is defined as
\begin{eqnarray} \label{eq:topological}
\chi_t=\frac{d^{2} \mathcal{V}(\theta,T,\mu)}{d \theta^{2}}\bigg|_{\theta=0},
\end{eqnarray}
where $\mathcal{V}(\theta,T,\mu)=\Omega(T,\mu,\sigma=\bar{\sigma},\eta=\bar{\eta})$ is the effective potential for the QCD $\theta$-vacuum at finite temperature and chemical potential.
Numerically,
the topological susceptibility estimated from Eq.~(\ref{eq:topological}) at zero temperature and chemical potential within the NJL model is~\cite{Lu-2018ukl}
\begin{eqnarray}
\chi_t^{1/4}=79.87~\mathrm{MeV},
\end{eqnarray}
which is in good agreement
with SU(2) CHPT prediction
$
\chi_t^{1/4}=77.8(4)~\mathrm{MeV}  
$~\cite{GrillidiCortona-2015jxo} and lattice simulations
$
\chi_t^{1/4}=78.1(2)~\mathrm{MeV}
$~\cite{16Borsanyi.others69-71N} in the isospin symmetric case.



In Fig.~\ref{fig:chiT}, we plot the topological susceptibility as a function of the temperature for $\mu=0,~120,~240,~360$ MeV (left panel) and of the chemical potential for $T=0,~50,~100,~200$ MeV (right panel).
One can find that the topological susceptibility in both panels decreases monotonically with
increasing temperature and/or chemical potential.
This suggests that the restoration of U(1)$_A$ symmetry is catalyzed by temperature and/or chemical potential.
This can be seen from the left panel of Fig.~\ref{fig:chiT} that for a larger chemical potential, the topological susceptibility starts to decrease earlier than that of the curve with a smaller chemical potential.
Furthermore, comparing the curves in Fig.~\ref{fig:chiT} with those in Fig.~\ref{fig:sigmaT}, one can clearly see that in each panel the curves with the same parameters show very similar behavior with increasing temperature (left panel) and chemical potential (right panel),
indicating that the evolution of the topological susceptibility is significantly affected by the chiral phase transition. However, there is an important difference between chiral condensate and topological susceptibility regarding high temperature and chemical potential limits: for the chiral condensate in both panels of Fig.~\ref{fig:sigmaT}, all curves converge to zero at high temperature and/or chemical potential, while the topological susceptibility shown in Fig.~\ref{fig:chiT} do not. This means that the U(1)$_A$ symmetry can still be considerably broken when the chiral symmetry is restored, which agrees with the previous study in the NJL model with different procedure~\cite{Fukushima-2001hr} and in other QCD effective QCD at finite temperature~\cite{16Jiang.Xia.ea74006-74006PRD,Kawaguchi-2023olk}.
Moreover, it is worth mentioning that our results are also in qualitative agreement with lattice simulations on the evolution of the topological susceptibility with the chemical potential~\cite{Iida-2019rah}.


\begin{figure*}[htb]  
  \includegraphics[width=0.48\textwidth]{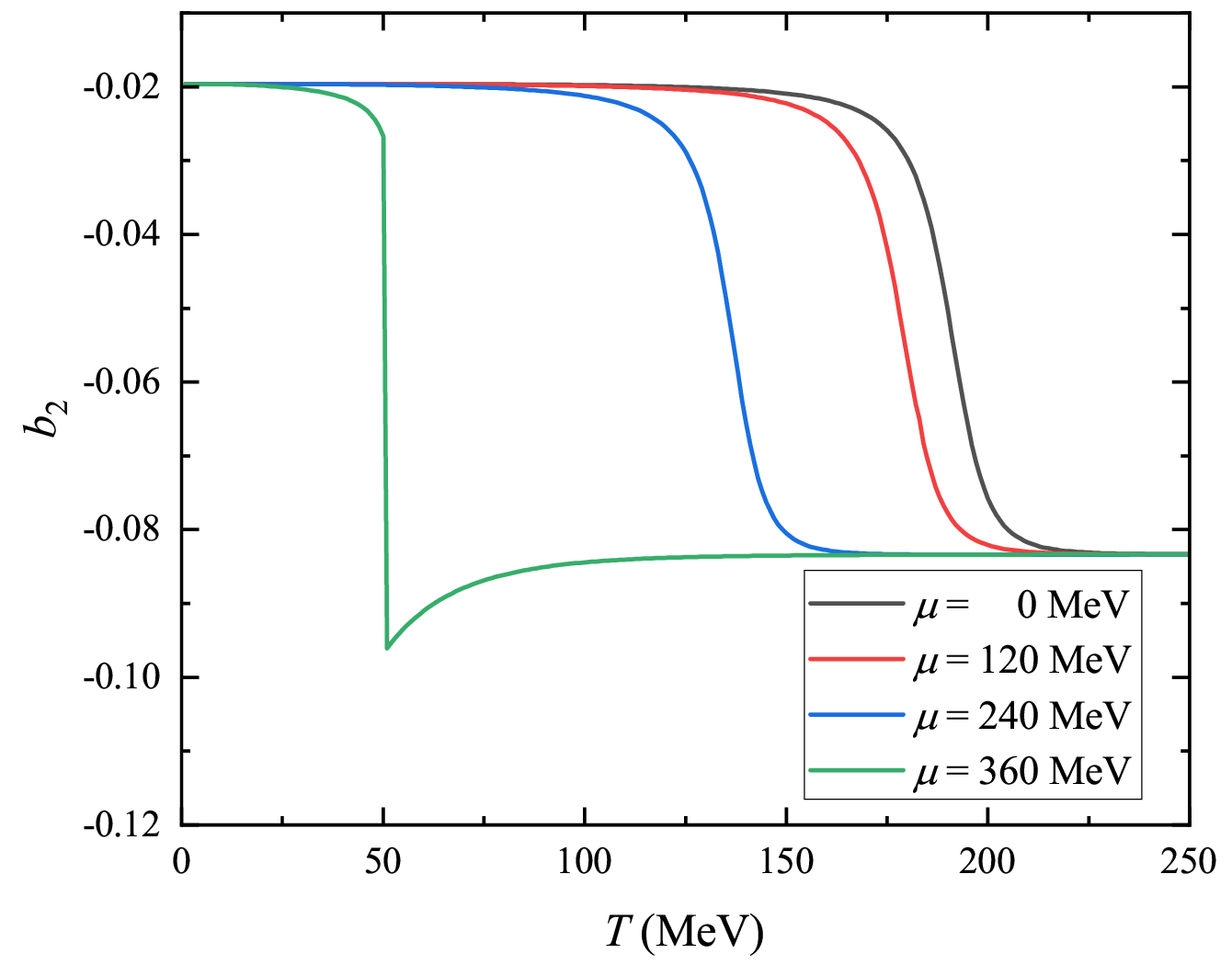}
   \includegraphics[width=0.48\textwidth]{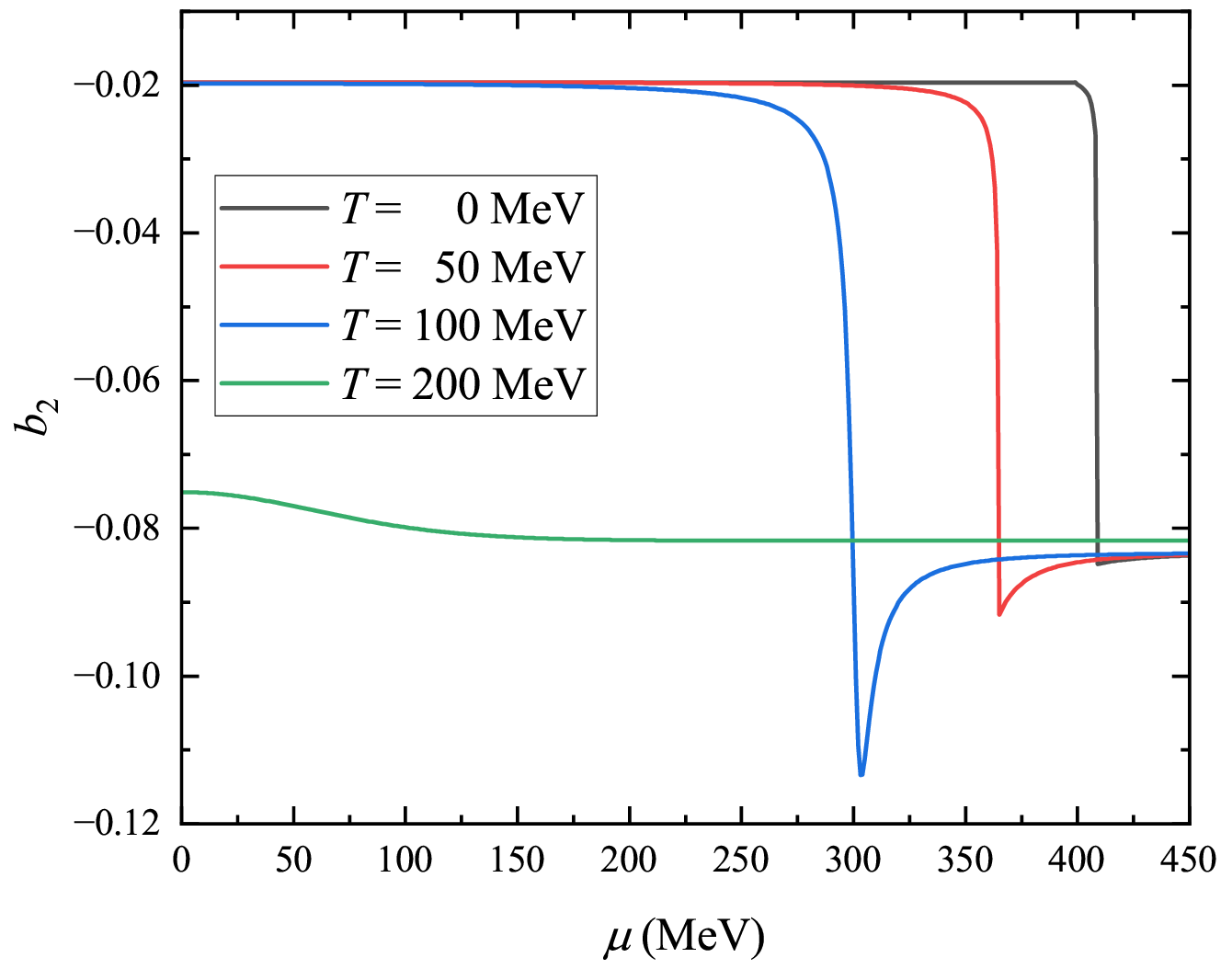}\\
  \caption{Variation of the normalized fourth cumulant, scaled by its value in the vacuum, with respect to the temperature at different chemical potentials
  (left panel) and to the chemical potential at different temperatures (right panel), respectively. Conventions for colors  are the same used in Fig.~\ref{fig:sigmaT}.
  }\label{fig:b2T}
\end{figure*}

As a by-product, we also provide an estimate of the topological quartic moment
of the topological charge, namely the normalized fourth cumulant~\cite{Vicari-2008jw,16Bonati.DElia.ea25028-25028PRD}
\begin{align} \label{eq:b2b2}
b_2 =\frac{1}{12\chi_t}\frac{d^{4} \mathcal{V}(\theta,T,\mu)}{d \theta^{4}}\bigg|_{\theta=0}
\end{align}
where
$\chi_t$ denotes the topological susceptibility defined in Eq.~(\ref{eq:topological}).
At zero temperature and chemical potential, the value of the normalized fourth cumulant computed in SU(2) CHPT up to next-to-leading order in the isospin limit is estimated to be $b_2(T=\mu=0)=-0.022(1)$~\cite{Bonati-2015vqz}, where the error is mainly due to the uncertainty on the coupling constants of CHPT~\cite{GrillidiCortona-2015jxo}.
The NJL model computation gives $b_2(T=\mu=0)=-0.020$,
which agrees perfectly well with the CHPT prediction.

In the left panel of Fig.~\ref{fig:b2T}, we show the variation of the normalized fourth cumulant $b_2$ with temperature for different chemical potentials $\mu=0,~120,~ 240,~360$ MeV.
In particular,
the green solid curve with $\mu = 360$ MeV remains practically constant up to the threshold around $T\simeq 50$ MeV, which corresponds to the critical temperature for the first order phase transition, and suddenly drops to a more negative value at this point. Beyond this point,
the normalized fourth cumulant
increases rapidly and soon converges to
the asymptotic value predicted by the dilute instanton gas model
~\cite{Bonati-2013tt,15Bonati6-6J},
i.e.,
\begin{eqnarray}
b_2^{\text {inst }}=-\frac{1}{12} \simeq-0.083.
\end{eqnarray}
The figure also shows the curves at
$\mu = 0,~120,~240$ MeV, represented by blue, red, and black solid curves respectively.
These three curves first remain unchanged at low temperatures and then decrease monotonically with the increase in temperature. The only difference is that the curve with a larger chemical potential tends to decrease at a lower temperature, indicating that the presence of the chemical potential shifts the critical point for the chiral phase transition to a lower temperature.
It is worth noting from Fig.~~\ref{fig:b2T} that the normalized fourth cumulant reaches a uniform value at high temperatures in all cases, regardless of the varying chemical potentials. This implies that the normalized fourth cumulant becomes insensitive to the QCD medium at sufficiently high temperatures and chemical potentials.

The right panel of Fig.~\ref{fig:b2T} shows the variation of the normalized fourth cumulant $b_2$ with respect to the temperature for different temperatures $T=0,~50,~100,~200$ MeV, represented by the solid black, red, blue, and green curves respectively. Apart from the green curve representing $T=200$ MeV, which shows a small change and is almost linearly dependent on the chemical potential,
the remaining three curves exhibit significant fluctuations as the chemical potential increases.
On the other hand, by increasing the temperature from $T=0$ MeV up to $T = 50$ MeV,
the discontinuity point is
pushed towards smaller values of chemical potential, which implies that the critical chemical potential decreases with the temperature. However, by further increasing the temperature up to $T=100$ MeV, as represented by the blue curve, the normalized fourth cumulant remains
continuous and finite with the change in the chemical potential.
This fact suggests
that by increasing the temperature from zero to a large enough temperature, the first order phase transition has translated into a chiral crossover.
Eventually, these curves all converge towards the same limit value as the chemical potential increases, which  means that the contribution from the temperature dependence of the terms shown in Eq.~(\ref{eq:OmegaQ}) vanishes when the chemical potential is large enough.

\subsection{Axion mass and its self-coupling constant}

We note that topological susceptibility $\chi_t$ is 
related to the axion potential as a function of temperature and chemical potential~\cite{Borsanyi-2015cka}. In particular, the axion mass is related to the topological susceptibility by~\cite{Berkowitz-2015aua,Petreczky-2016vrs,Bonati-2018blm,Gorghetto-2018ocs,Horvatic-2019eok}
\begin{eqnarray} \label{eq:ma2}
m_a^2=\left.\frac{\mathrm{d}^2 \mathcal{V}(a,T,\mu)}{\mathrm{d} a^2}\right|_{a=0}=\left.\frac{\mathrm{d}^2 \mathcal{V}(\theta,T,\mu)}{f_a^2\mathrm{d} \theta^2}\right|_{\theta=0}=\frac{\chi_t}{f_a^2},
\end{eqnarray}
where $f_a$ and $\chi_t$ are the axion decay constant and topological susceptibility, respectively.
In the above equation, we have set $\theta=a/f_a$ based on the PQ mechanism and the relation between the effective potential of $\theta$-vacuum
and axion potential~\cite{Lu-2018ukl,Lu-2020rhp}.
From Eq.~(\ref{eq:ma2}),
the axion mass in the vacuum within the NJL model is~\cite{Lu-2018ukl}
$
m_{a0}=6.38\times10^3/f_a~\mathrm{MeV},
$
in agreement with the result of CHPT in the isospin symmetric
case~\cite{GrillidiCortona-2015jxo}, $m_{a0}=6.06(5)\times10^3/f_a~\mathrm{MeV}$,
as well as with that of the invisible axion model, $m_{a0}\simeq6.00\times10^3/f_a~\mathrm{MeV}$~\cite{87Kim1-177PR,Cheng-1987gp,90Turner67-97PR,90Raffelt1-113PR}.

Despite the agreement in zero temperature results for the axion mass from various methods, it is essential to emphasize that the precise value of the axion mass has not been well determined so far, primarily because of the lack of knowledge regarding the axion decay constant $f_a$.
For the invisible axion~\cite{81Dine.Fischler.ea199-202PLB,Zhitnitsky-1980tq,Kim-1979if,Shifman-1979if}, the axion mass window is typically from about $10^{-6}$~eV to $10^{-2}$~eV.
According to the constraints from astrophysical observations, the so-called classical axion window is
$
10^9~\mathrm{GeV}\lesssim f_a\lesssim10^{12}~\mathrm{GeV}
$~\cite{16Marsh1-79PR,DiLuzio-2020wdo},
with the upper and lower bounds set by the observed abundance of dark
matter and the neutrino burst duration of SN1987A, respectively. However, invisible axion models with $f_a\gtrsim 10^{8}$ GeV suffer from the axion quality problem~\cite{Georgi-1981pu,Kamionkowski-1992mf,Holman-1992us}.
Consequently, in order to solve the axion quality problem and strong CP problem simultaneously, the axions with a heavy mass and low decay constant $f_a$ have been proposed~\cite{Rubakov-1997vp,Berezhiani-2000gh}. Furthermore, after revisiting the constraints on the MeV mass window for the QCD axion and its variants, the authors in Ref.~\cite{Alves-2017avw} claimed that there is still a possibility for a viable QCD axion model with a mass in the MeV range. Nevertheless, these aspects are not the main focus of our research, as we are specifically interested in exploring how the properties of axions vary in different environments. In the following, we will restrict our analysis to the low-energy properties of axions in a dense and hot QCD medium.

\begin{figure*}[htb]  
  \includegraphics[width=0.48\textwidth]{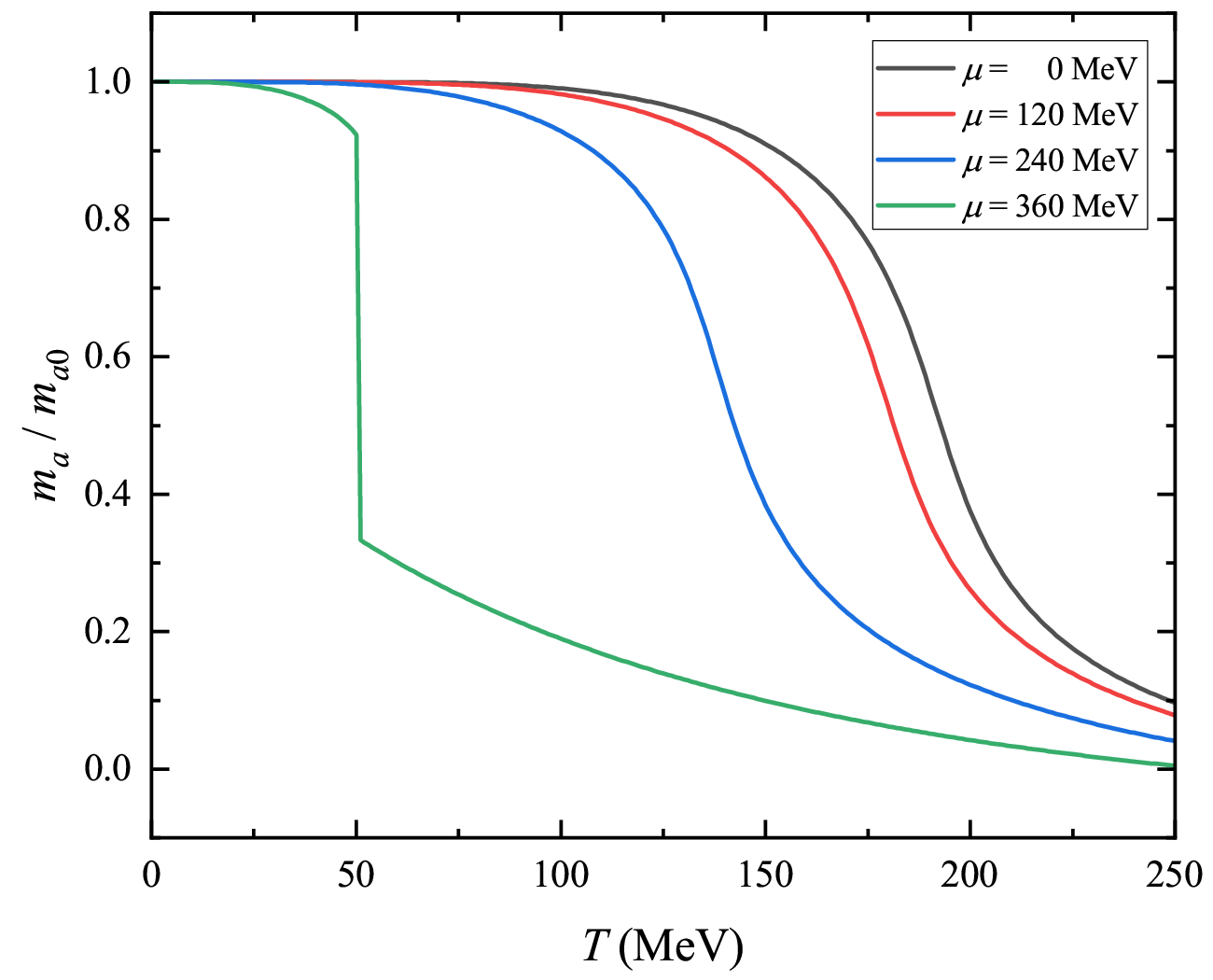}
   \includegraphics[width=0.48\textwidth]{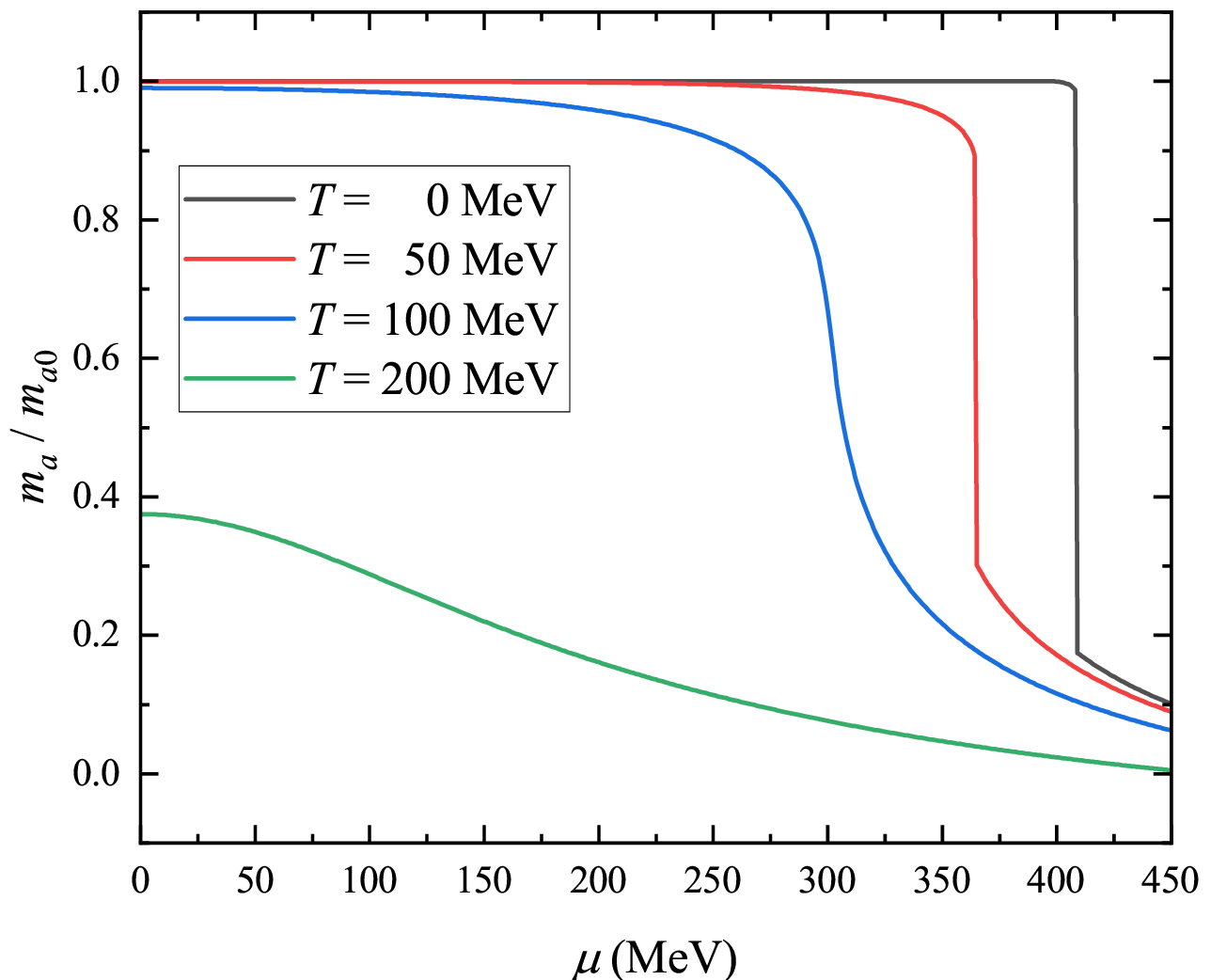}\\
  \caption{Variation of the axion mass, scaled by its value in the vacuum, with respect to the temperature at different chemical potentials
  (left panel) and to the chemical potential at different temperatures  (right panel), respectively. Conventions for colors  are the same used in Fig.~\ref{fig:sigmaT}.
  }\label{fig:mafaT}
\end{figure*}

As shown in Fig.~\ref{fig:mafaT}, we plot the axion mass, scaled by its value in the vacuum, as a function of the temperature for $\mu=0,~120,~240,~360$ MeV (left panel) and of the chemical potential for $T=0,~50,~100,~200$ MeV (right panel).
These results have been obtained by using Eq.~(\ref{eq:ma2}) with
the solution of the gap equation from Eq.~(\ref{eq:gapEquations2}).
Both panels show a decrease in axion mass with higher temperatures and/or chemical potentials, indicating that considering temperature and finite density effects tends to lower the axion mass.
Moreover, by comparing the axion mass depicted in this figure with the results for the chiral condensate and the topological susceptibility shown in Fig.~\ref{fig:sigmaT} and Fig.~\ref{fig:chiT}, one can observe that the evolution of the axion mass with respect to temperature (left panel) and chemical potential (right panel) follows a similar behavior as that of the chiral condensate and the topological susceptibility.
This can be well understood that the behavior of axion mass follows the response of topological susceptibility to the temperature~\cite{16Borsanyi.others69-71N} and is determined precisely by Eq.~(\ref{eq:ma2}).
Furthermore, the results for the behavior of the axion mass with respect to the chemical potential agree with the CHPT calculations at vanishing temperature~\cite{Balkin-2020dsr}.
However, as stated in Ref.~\cite{Zhang-2023lij}, the results from CHPT at considerable chemical potential and high temperature cannot be taken seriously, since it lacks information
about the QCD phase transitions in this regime.

\begin{figure*}[htb]  
  \includegraphics[width=0.48\textwidth]{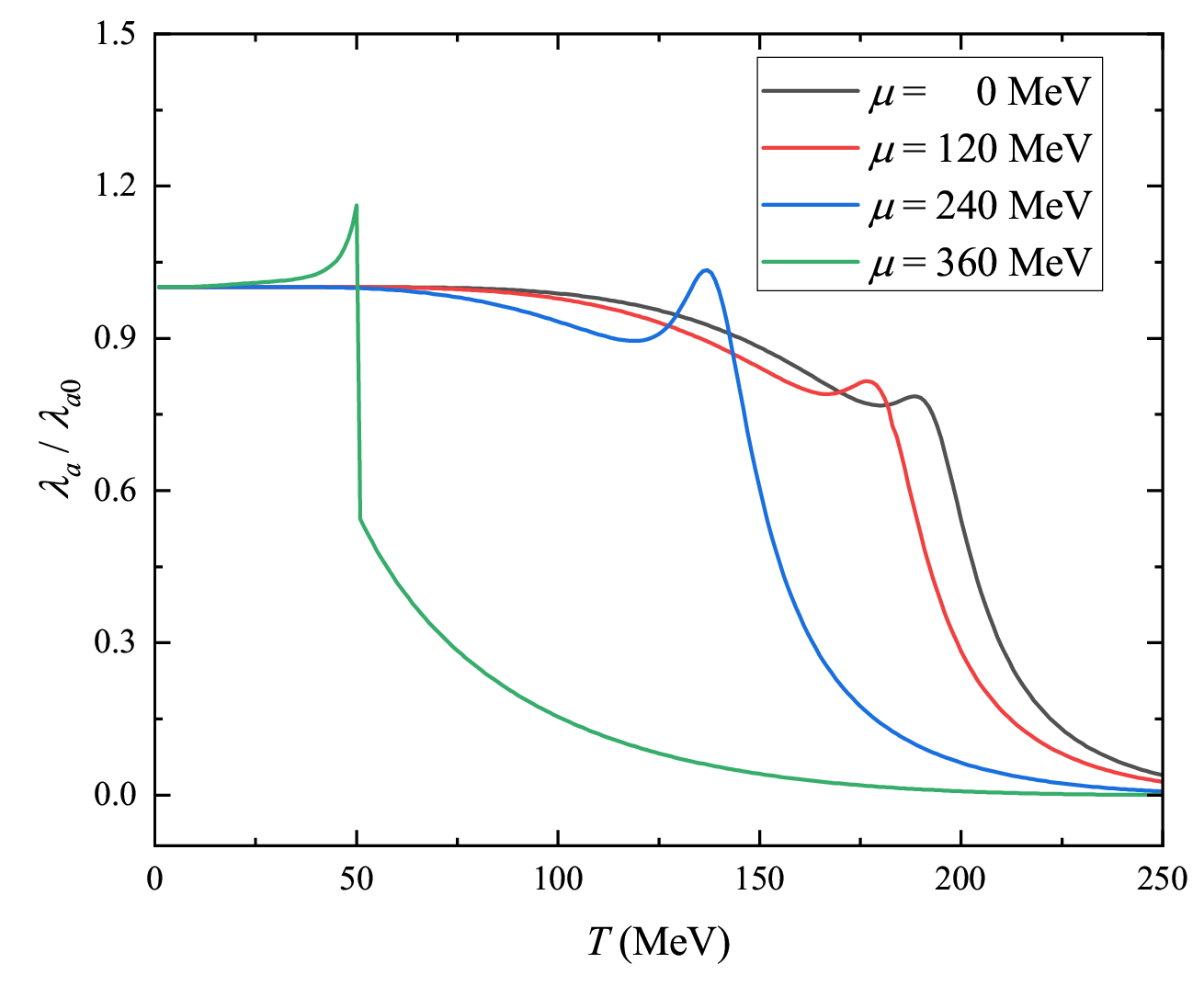}
   \includegraphics[width=0.48\textwidth]{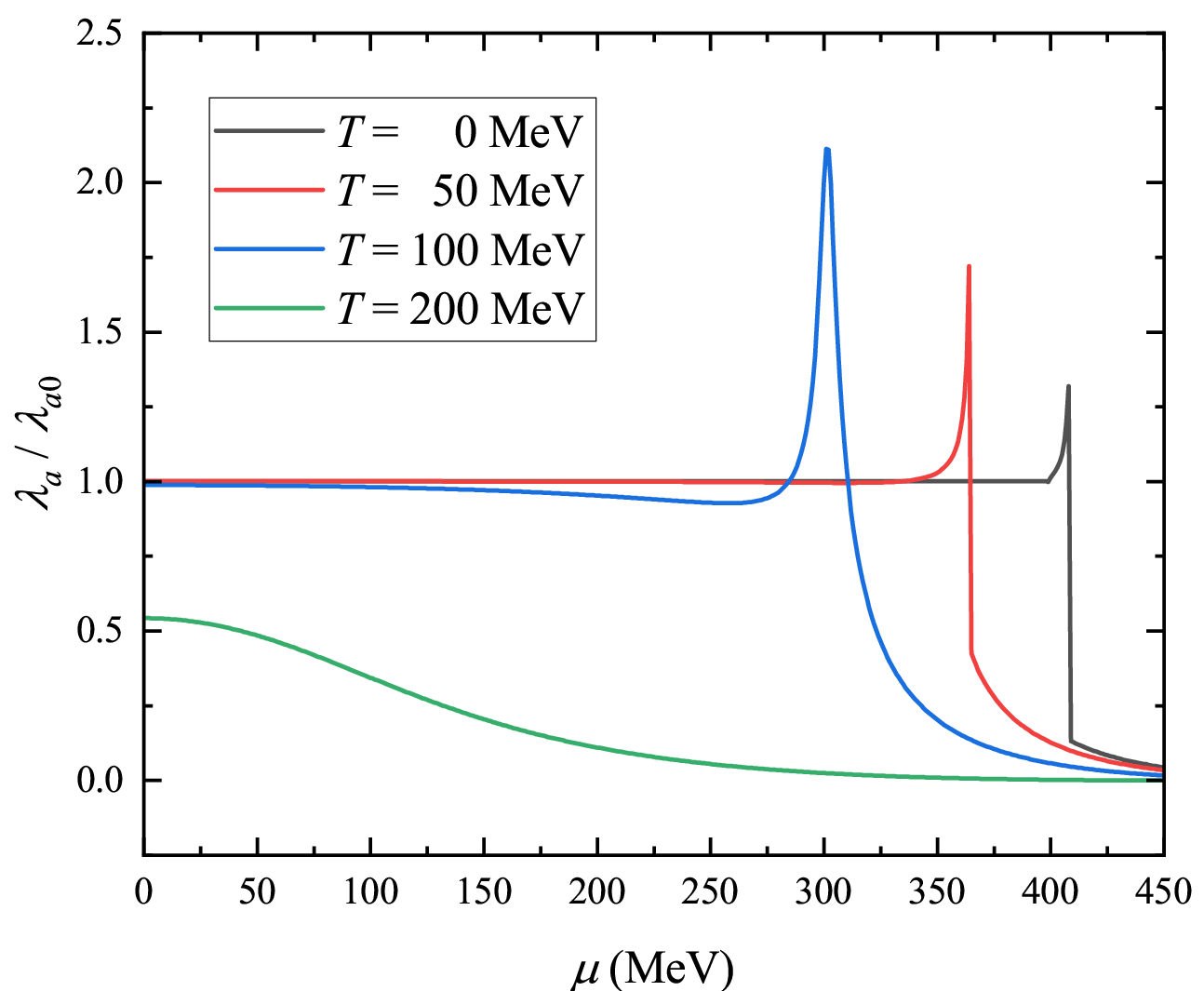}\\
  \caption{Variation of the axion self-coupling constant, scaled by its value in the vacuum, with respect to the temperature at different chemical potentials
  (left panel) and to the chemical potential at different temperatures
  (right panel), respectively. Conventions for colors  are the same used in Fig.~\ref{fig:sigmaT}.
  }\label{fig:lambdaAT}
\end{figure*}

The axion self-coupling plays an important role in the formation of a Bose-Einstein condensate (BEC)~\cite{Sikivie-2009qn} and the so-called axion stars~\cite{Tkachev-1986tr,Kolb-1993zz,Hogan-1988mp,11Barranco.Bernal43525-43525PRD,11Chavanis43531-43531PRD,Zhang-2018slz}.
Similar to the definition of the axion mass in Eq.~(\ref{eq:ma2}), the axion self-coupling
is determined by the fourth derivative of the
axion potential
at the point where the axion field equals zero, namely
\begin{eqnarray}
\lambda_a=\left.\frac{\mathrm{d}^4 \mathcal{V}(a,T,\mu)}{\mathrm{d} a^4}\right|_{a=0} .
\end{eqnarray}
At zero temperature and chemical potential, this  can be computed within the NJL model, namely,
$
\lambda_{a0}=-(55.64~\mathrm{MeV})^4/f_a^4,
$
in quantitative agreement with the CHPT prediction, $\lambda_{a0}=-(55.79(92)~\mathrm{MeV}/f_a)^4$~\cite{GrillidiCortona-2015jxo} in the case of two degenerate quark masses.

Fig.~\ref{fig:lambdaAT}, on the left panel, displays the variation of the normalized self-coupling with temperature.
It is obvious that as the temperature increases,
each curve forms a peak at the critical point, and after that, the normalized self-coupling steadily decreases as the temperature continues to rise.
In particular, by increasing the chemical potential from zero to a larger value,
the peak structure along the temperature axis goes sharper and shifts to lower temperatures. However, if the given chemical potential is large enough, such as the green curve at $\mu=360$ MeV, the peak evolves into a singularity, signaling a first order phase transition.

In the right panel of Fig.~\ref{fig:lambdaAT}, we display
the axion self-coupling constant, scaled by its value in the vacuum, as a function of the chemical potential for $T=0,~50,~100,~200$ MeV.
Similar to the behavior of the axion mass at low chemical potentials, such as the black curve at $T=0$ MeV shown in the right panel of Fig.~\ref{fig:mafaT}, the axion self-coupling constant remains almost constant until the critical point around $\mu\simeq 408$ MeV. However, the axion mass and self-coupling constant show quite different behavior around the critical point: the former drops sharply to a lower value at the critical point, while the latter, on the contrary, first raises to a larger value, and then drops to a lower value beyond this critical point. The magnitude of the change at the critical point is significant, and it can be more than twice the
vacuum value. This suggests that the chemical potential plays a crucial role in increasing the self-interaction of the axion, particularly in response to the chiral phase transition at different temperatures and chemical potentials. Whereas, for the blue curve at $T=200$ MeV, at which the system is in the chiral restoration phase the self-coupling constant starts at $0.55\lambda_{a0}$ and decreases monotonically with increasing chemical potential in the considered chemical potential range.

We close this section with a remark on the potential for the creation or enhancement of an axion BEC within the compact stars. Dense objects like neutron stars and hybrid stars can generate and trap axions and subsequently participate in energy transport that affects their thermal evolution~\cite{Raffelt-2006cw,Sedrakian-2015krq}.
In this case, compact stars could contain a considerable number of axions, particularly the axion dark matter~\cite{Bramante-2023djs,Karkevandi-2024vov}. We hence conjecture that in such a high baryonic density environment, a significant increase of axion self-interaction may occur, which will then lead to the formation or further enhancement of the axion BEC inside the compact stars.
In fact, recently the presence of axions are found to have a non-negligible effects on the structure, maximum mass and tidal
deformability of hybrid and neutron stars~\cite{Lopes-2022efy,Karkevandi-2021ygv}.

\section{Conclusion}  \label{sec:CONCLUSION}

We have studied the effects of temperature and chemical potential on the first two lowest cumulants of the QCD topological charge distribution, namely the topological susceptibility and the normalized fourth cumulant, as well as on the axion properties. The QCD medium at finite temperature and chemical potential has been described by the NJL model with two flavors, while the axion field is incorporated into the NJL Lagrangian through the U(1)$_A$ symmetry breaking term by employing the Peccei-Quinn mechanism. We found that the chiral phase transition is a first order phase transition at low temperature and large chemical potential, and it becomes a chiral crossover at high temperature and low chemical potential, signaled by the discontinuity and smooth behavior of the chiral condensate, respectively.

By comparing the zero temperature results for the topological susceptibility, normalized fourth cumulant, axion mass, and axion self-coupling constant obtained in the present work with those obtained in CHPT and first-principle lattice simulations, we found that they are in good agreement with each other.
Moreover, we found that the chiral phase transition and QCD crossover have significant effects on the behavior of the QCD topology and the axion properties in a hot medium.
In particular, the evolution of the topological susceptibility and the axion mass with temperature and/or chemical potential shows similar behavior as that of the chiral condensate.
In conclusion, the chiral phase transition significantly reduces the axion mass while considerably enhancing the self-coupling constant, which could lead to the formation or further enhancement of an axion BEC in compact astrophysical objects with a dense QCD medium at zero or low temperatures. Yet, we should emphasize that a more quantitative study is necessary for substantiating this conjecture.


Recent studies have suggested the possibility of additional terms in the anomaly sector~\cite{Pisarski:2024esv,Pisarski-2019upw}. If true, this could imply the existence of another way to couple the axion to quarks and thus modify the axion dynamics. For example, the axion decays into a photon via a loop of quarks~\cite{16Marsh1-79PR,Aghaie:2024jkj}, a process that is phenomenologically important. The discussion of this process is crucial as it provides insights into the interactions between axions and quarks, potentially reshaping our understanding of axion physics.
A step beyond, which deserves being explored
in more detail,
to extend our current work by delving into the isospin density effect~\cite{Lu-2019diy,Lu-2021hvw} and studying the influence of magnetic fields~\cite{Lu-2022khf}, both at zero temperature and under varying thermal conditions.
By incorporating these additional variables, we
expect to broaden
the scope of our investigations and advance our comprehension of the underlying mechanisms governing these phenomena. We plan to report on these subjects in the near future.


\section*{Acknowledgments}

The authors thank Marco Ruggieri for useful discussions.
ZYL is supported by
the National Natural Science Foundation of China
(Grant Nos.~12205093 and 11835015),
the Hunan Provincial Natural Science Foundation of China (Grant No.~2021JJ40188),
and the Scientific Research Fund of Hunan Provincial Education Department of China (Grant No.~19C0772). This work is also supported in part by the National Natural Science Foundation of China (Grant Nos.~12375045, 12404240 and 12204166), and by the China Postdoctoral Science Foundation (Grant No.~2024M750489).

\bibliographystyle{aapmrev4-2}  
\bibliography{RefLuInsp}

\begin{thebibliography}{133}%
\makeatletter
\providecommand \@ifxundefined [1]{%
 \@ifx{#1\undefined}
}%
\providecommand \@ifnum [1]{%
 \ifnum #1\expandafter \@firstoftwo
 \else \expandafter \@secondoftwo
 \fi
}%
\providecommand \@ifx [1]{%
 \ifx #1\expandafter \@firstoftwo
 \else \expandafter \@secondoftwo
 \fi
}%
\providecommand \natexlab [1]{#1}%
\providecommand \enquote  [1]{``#1''}%
\providecommand \bibnamefont  [1]{#1}%
\providecommand \bibfnamefont [1]{#1}%
\providecommand \citenamefont [1]{#1}%
\providecommand \href@noop [0]{\@secondoftwo}%
\providecommand \href [0]{\begingroup \@sanitize@url \@href}%
\providecommand \@href[1]{\@@startlink{#1}\@@href}%
\providecommand \@@href[1]{\endgroup#1\@@endlink}%
\providecommand \@sanitize@url [0]{\catcode `\\12\catcode `\$12\catcode
  `\&12\catcode `\#12\catcode `\^12\catcode `\_12\catcode `\%12\relax}%
\providecommand \@@startlink[1]{}%
\providecommand \@@endlink[0]{}%
\providecommand \url  [0]{\begingroup\@sanitize@url \@url }%
\providecommand \@url [1]{\endgroup\@href {#1}{\urlprefix }}%
\providecommand \urlprefix  [0]{URL }%
\providecommand \Eprint [0]{\href }%
\providecommand \doibase [0]{https://doi.org/}%
\providecommand \selectlanguage [0]{\@gobble}%
\providecommand \bibinfo  [0]{\@secondoftwo}%
\providecommand \bibfield  [0]{\@secondoftwo}%
\providecommand \translation [1]{[#1]}%
\providecommand \BibitemOpen [0]{}%
\providecommand \bibitemStop [0]{}%
\providecommand \bibitemNoStop [0]{.\EOS\space}%
\providecommand \EOS [0]{\spacefactor3000\relax}%
\providecommand \BibitemShut  [1]{\csname bibitem#1\endcsname}%
\let\auto@bib@innerbib\@empty
\bibitem [{\citenamefont {Gross}\ \emph {et~al.}(2023)\citenamefont {Gross}
  \emph {et~al.}}]{Gross-2022hyw}%
  \BibitemOpen
  \bibfield  {author} {\bibinfo {author} {\bibfnamefont {F.}~\bibnamefont
  {Gross}} \emph {et~al.},\ }\href
  {https://doi.org/10.1140/epjc/s10052-023-11949-2} {\bibfield  {journal}
  {\bibinfo  {journal} {Eur. Phys. J. C}\ }\textbf {\bibinfo {volume} {83}},\
  \bibinfo {pages} {1125} (\bibinfo {year} {2023})},\ \Eprint
  {https://arxiv.org/abs/2212.11107} {arXiv:2212.11107 [hep-ph]} \BibitemShut
  {NoStop}%
\bibitem [{\citenamefont {Gross}, \citenamefont {Pisarski},\ and\ \citenamefont
  {Yaffe}(1981)}]{81Gross.Pisarski.ea43-43RMP}%
  \BibitemOpen
  \bibfield  {author} {\bibinfo {author} {\bibfnamefont {D.~J.}\ \bibnamefont
  {Gross}}, \bibinfo {author} {\bibfnamefont {R.~D.}\ \bibnamefont
  {Pisarski}},\ and\ \bibinfo {author} {\bibfnamefont {L.~G.}\ \bibnamefont
  {Yaffe}},\ }\href {https://doi.org/10.1103/RevModPhys.53.43} {\bibfield
  {journal} {\bibinfo  {journal} {Rev. Mod. Phys.}\ }\textbf {\bibinfo {volume}
  {53}},\ \bibinfo {pages} {43} (\bibinfo {year} {1981})}\BibitemShut {NoStop}%
\bibitem [{\citenamefont {'t~Hooft}(1986)}]{86Hooft357-387PR}%
  \BibitemOpen
  \bibfield  {author} {\bibinfo {author} {\bibfnamefont {G.}~\bibnamefont
  {'t~Hooft}},\ }\href {https://doi.org/10.1016/0370-1573(86)90117-1}
  {\bibfield  {journal} {\bibinfo  {journal} {Phys. Rep.}\ }\textbf {\bibinfo
  {volume} {142}},\ \bibinfo {pages} {357} (\bibinfo {year}
  {1986})}\BibitemShut {NoStop}%
\bibitem [{\citenamefont {Sch\"afer}\ and\ \citenamefont
  {Shuryak}(1998)}]{98Schdotafer.Shuryak323-426RMP}%
  \BibitemOpen
  \bibfield  {author} {\bibinfo {author} {\bibfnamefont {T.}~\bibnamefont
  {Sch\"afer}}\ and\ \bibinfo {author} {\bibfnamefont {E.~V.}\ \bibnamefont
  {Shuryak}},\ }\href {https://doi.org/10.1103/RevModPhys.70.323} {\bibfield
  {journal} {\bibinfo  {journal} {Rev. Mod. Phys.}\ }\textbf {\bibinfo {volume}
  {70}},\ \bibinfo {pages} {323} (\bibinfo {year} {1998})},\ \Eprint
  {https://arxiv.org/abs/hep-ph/9610451} {arXiv:hep-ph/9610451 [hep-ph]}
  \BibitemShut {NoStop}%
\bibitem [{\citenamefont {'t~Hooft}(1976{\natexlab{a}})}]{76Hooft8-11PRL}%
  \BibitemOpen
  \bibfield  {author} {\bibinfo {author} {\bibfnamefont {G.}~\bibnamefont
  {'t~Hooft}},\ }\href {https://doi.org/10.1103/PhysRevLett.37.8} {\bibfield
  {journal} {\bibinfo  {journal} {Phys. Rev. Lett.}\ }\textbf {\bibinfo
  {volume} {37}},\ \bibinfo {pages} {8} (\bibinfo {year}
  {1976}{\natexlab{a}})},\ \bibinfo {note} {[,226(1976)]}\BibitemShut {NoStop}%
\bibitem [{\citenamefont {'t~Hooft}(1976{\natexlab{b}})}]{76Hooft3432-3450PRD}%
  \BibitemOpen
  \bibfield  {author} {\bibinfo {author} {\bibfnamefont {G.}~\bibnamefont
  {'t~Hooft}},\ }\href {https://doi.org/10.1103/PhysRevD.18.2199.3,
  10.1103/PhysRevD.14.3432} {\bibfield  {journal} {\bibinfo  {journal} {Phys.
  Rev. D}\ }\textbf {\bibinfo {volume} {14}},\ \bibinfo {pages} {3432}
  (\bibinfo {year} {1976}{\natexlab{b}})}\BibitemShut {NoStop}%
\bibitem [{\citenamefont {Guo}\ and\ \citenamefont
  {Mei\ss{}ner}(2015)}]{Guo-2015oxa}%
  \BibitemOpen
  \bibfield  {author} {\bibinfo {author} {\bibfnamefont {F.-K.}\ \bibnamefont
  {Guo}}\ and\ \bibinfo {author} {\bibfnamefont {U.-G.}\ \bibnamefont
  {Mei\ss{}ner}},\ }\href {https://doi.org/10.1016/j.physletb.2015.07.076}
  {\bibfield  {journal} {\bibinfo  {journal} {Phys. Lett. B}\ }\textbf
  {\bibinfo {volume} {749}},\ \bibinfo {pages} {278} (\bibinfo {year}
  {2015})},\ \Eprint {https://arxiv.org/abs/1506.05487} {arXiv:1506.05487
  [hep-ph]} \BibitemShut {NoStop}%
\bibitem [{\citenamefont {Kawaguchi}\ and\ \citenamefont
  {Suenaga}(2023)}]{Kawaguchi-2023olk}%
  \BibitemOpen
  \bibfield  {author} {\bibinfo {author} {\bibfnamefont {M.}~\bibnamefont
  {Kawaguchi}}\ and\ \bibinfo {author} {\bibfnamefont {D.}~\bibnamefont
  {Suenaga}},\ }\href {https://doi.org/10.1007/JHEP08(2023)189} {\bibfield
  {journal} {\bibinfo  {journal} {JHEP}\ }\textbf {\bibinfo {volume} {08}},\
  \bibinfo {pages} {189} (\bibinfo {year} {2023})},\ \Eprint
  {https://arxiv.org/abs/2305.18682} {arXiv:2305.18682 [hep-ph]} \BibitemShut
  {NoStop}%
\bibitem [{\citenamefont {Crewther}\ \emph {et~al.}(1979)\citenamefont
  {Crewther}, \citenamefont {Di~Vecchia}, \citenamefont {Veneziano},\ and\
  \citenamefont {Witten}}]{79Crewther.DiVecchia.ea123-123PLB}%
  \BibitemOpen
  \bibfield  {author} {\bibinfo {author} {\bibfnamefont {R.~J.}\ \bibnamefont
  {Crewther}}, \bibinfo {author} {\bibfnamefont {P.}~\bibnamefont
  {Di~Vecchia}}, \bibinfo {author} {\bibfnamefont {G.}~\bibnamefont
  {Veneziano}},\ and\ \bibinfo {author} {\bibfnamefont {E.}~\bibnamefont
  {Witten}},\ }\href {https://doi.org/10.1016/0370-2693(80)91025-4,
  10.1016/0370-2693(79)90128-X} {\bibfield  {journal} {\bibinfo  {journal}
  {Phys. Lett. B}\ }\textbf {\bibinfo {volume} {88}},\ \bibinfo {pages} {123}
  (\bibinfo {year} {1979})},\ \bibinfo {note} {[Erratum: Phys.
  Lett.91B,487(1980)]}\BibitemShut {NoStop}%
\bibitem [{\citenamefont {Baker}\ \emph {et~al.}(2006)\citenamefont {Baker}
  \emph {et~al.}}]{Baker-2006ts}%
  \BibitemOpen
  \bibfield  {author} {\bibinfo {author} {\bibfnamefont {C.~A.}\ \bibnamefont
  {Baker}} \emph {et~al.},\ }\href
  {https://doi.org/10.1103/PhysRevLett.97.131801} {\bibfield  {journal}
  {\bibinfo  {journal} {Phys. Rev. Lett.}\ }\textbf {\bibinfo {volume} {97}},\
  \bibinfo {pages} {131801} (\bibinfo {year} {2006})},\ \Eprint
  {https://arxiv.org/abs/hep-ex/0602020} {arXiv:hep-ex/0602020 [hep-ex]}
  \BibitemShut {NoStop}%
\bibitem [{\citenamefont {Griffith}\ \emph {et~al.}(2009)\citenamefont
  {Griffith}, \citenamefont {Swallows}, \citenamefont {Loftus}, \citenamefont
  {Romalis}, \citenamefont {Heckel},\ and\ \citenamefont
  {Fortson}}]{09Griffith.Swallows.ea101601-101601PRL}%
  \BibitemOpen
  \bibfield  {author} {\bibinfo {author} {\bibfnamefont {W.~C.}\ \bibnamefont
  {Griffith}}, \bibinfo {author} {\bibfnamefont {M.~D.}\ \bibnamefont
  {Swallows}}, \bibinfo {author} {\bibfnamefont {T.~H.}\ \bibnamefont
  {Loftus}}, \bibinfo {author} {\bibfnamefont {M.~V.}\ \bibnamefont {Romalis}},
  \bibinfo {author} {\bibfnamefont {B.~R.}\ \bibnamefont {Heckel}},\ and\
  \bibinfo {author} {\bibfnamefont {E.~N.}\ \bibnamefont {Fortson}},\ }\href
  {https://doi.org/10.1103/PhysRevLett.102.101601} {\bibfield  {journal}
  {\bibinfo  {journal} {Phys. Rev. Lett.}\ }\textbf {\bibinfo {volume} {102}},\
  \bibinfo {pages} {101601} (\bibinfo {year} {2009})}\BibitemShut {NoStop}%
\bibitem [{\citenamefont {Parker}\ \emph {et~al.}(2015)\citenamefont {Parker}
  \emph {et~al.}}]{15Parker.others233002-233002PRL}%
  \BibitemOpen
  \bibfield  {author} {\bibinfo {author} {\bibfnamefont {R.~H.}\ \bibnamefont
  {Parker}} \emph {et~al.},\ }\href
  {https://doi.org/10.1103/PhysRevLett.114.233002} {\bibfield  {journal}
  {\bibinfo  {journal} {Phys. Rev. Lett.}\ }\textbf {\bibinfo {volume} {114}},\
  \bibinfo {pages} {233002} (\bibinfo {year} {2015})},\ \Eprint
  {https://arxiv.org/abs/1504.07477} {arXiv:1504.07477 [nucl-ex]} \BibitemShut
  {NoStop}%
\bibitem [{\citenamefont {Graner}\ \emph {et~al.}(2016)\citenamefont {Graner},
  \citenamefont {Chen}, \citenamefont {Lindahl},\ and\ \citenamefont
  {Heckel}}]{16Graner.Chen.ea161601-161601PRL}%
  \BibitemOpen
  \bibfield  {author} {\bibinfo {author} {\bibfnamefont {B.}~\bibnamefont
  {Graner}}, \bibinfo {author} {\bibfnamefont {Y.}~\bibnamefont {Chen}},
  \bibinfo {author} {\bibfnamefont {E.~G.}\ \bibnamefont {Lindahl}},\ and\
  \bibinfo {author} {\bibfnamefont {B.~R.}\ \bibnamefont {Heckel}},\ }\href
  {https://doi.org/10.1103/PhysRevLett.119.119901,
  10.1103/PhysRevLett.116.161601} {\bibfield  {journal} {\bibinfo  {journal}
  {Phys. Rev. Lett.}\ }\textbf {\bibinfo {volume} {116}},\ \bibinfo {pages}
  {161601} (\bibinfo {year} {2016})},\ \bibinfo {note} {[Erratum: Phys. Rev.
  Lett.119,no.11,119901(2017)]},\ \Eprint {https://arxiv.org/abs/1601.04339}
  {arXiv:1601.04339 [physics.atom-ph]} \BibitemShut {NoStop}%
\bibitem [{\citenamefont {Guo}\ \emph {et~al.}(2015)\citenamefont {Guo},
  \citenamefont {Horsley}, \citenamefont {Mei\ss{}ner}, \citenamefont
  {Nakamura}, \citenamefont {Perlt}, \citenamefont {Rakow}, \citenamefont
  {Schierholz}, \citenamefont {Schiller},\ and\ \citenamefont
  {Zanotti}}]{Guo-2015tla}%
  \BibitemOpen
  \bibfield  {author} {\bibinfo {author} {\bibfnamefont {F.~K.}\ \bibnamefont
  {Guo}}, \bibinfo {author} {\bibfnamefont {R.}~\bibnamefont {Horsley}},
  \bibinfo {author} {\bibfnamefont {U.~G.}\ \bibnamefont {Mei\ss{}ner}},
  \bibinfo {author} {\bibfnamefont {Y.}~\bibnamefont {Nakamura}}, \bibinfo
  {author} {\bibfnamefont {H.}~\bibnamefont {Perlt}}, \bibinfo {author}
  {\bibfnamefont {P.~E.~L.}\ \bibnamefont {Rakow}}, \bibinfo {author}
  {\bibfnamefont {G.}~\bibnamefont {Schierholz}}, \bibinfo {author}
  {\bibfnamefont {A.}~\bibnamefont {Schiller}},\ and\ \bibinfo {author}
  {\bibfnamefont {J.~M.}\ \bibnamefont {Zanotti}},\ }\href
  {https://doi.org/10.1103/PhysRevLett.115.062001} {\bibfield  {journal}
  {\bibinfo  {journal} {Phys. Rev. Lett.}\ }\textbf {\bibinfo {volume} {115}},\
  \bibinfo {pages} {062001} (\bibinfo {year} {2015})},\ \Eprint
  {https://arxiv.org/abs/1502.02295} {arXiv:1502.02295 [hep-lat]} \BibitemShut
  {NoStop}%
\bibitem [{\citenamefont {Peccei}\ and\ \citenamefont
  {Quinn}(1977{\natexlab{a}})}]{77Peccei.Quinn1440-1443PRL}%
  \BibitemOpen
  \bibfield  {author} {\bibinfo {author} {\bibfnamefont {R.~D.}\ \bibnamefont
  {Peccei}}\ and\ \bibinfo {author} {\bibfnamefont {H.~R.}\ \bibnamefont
  {Quinn}},\ }\href {https://doi.org/10.1103/PhysRevLett.38.1440} {\bibfield
  {journal} {\bibinfo  {journal} {Phys. Rev. Lett.}\ }\textbf {\bibinfo
  {volume} {38}},\ \bibinfo {pages} {1440} (\bibinfo {year}
  {1977}{\natexlab{a}})}\BibitemShut {NoStop}%
\bibitem [{\citenamefont {Peccei}\ and\ \citenamefont
  {Quinn}(1977{\natexlab{b}})}]{77Peccei.Quinn1791-1797PRD}%
  \BibitemOpen
  \bibfield  {author} {\bibinfo {author} {\bibfnamefont {R.~D.}\ \bibnamefont
  {Peccei}}\ and\ \bibinfo {author} {\bibfnamefont {H.~R.}\ \bibnamefont
  {Quinn}},\ }\href {https://doi.org/10.1103/PhysRevD.16.1791} {\bibfield
  {journal} {\bibinfo  {journal} {Phys. Rev. D}\ }\textbf {\bibinfo {volume}
  {16}},\ \bibinfo {pages} {1791} (\bibinfo {year}
  {1977}{\natexlab{b}})}\BibitemShut {NoStop}%
\bibitem [{\citenamefont {Weinberg}(1978)}]{78Weinberg223-226PRL}%
  \BibitemOpen
  \bibfield  {author} {\bibinfo {author} {\bibfnamefont {S.}~\bibnamefont
  {Weinberg}},\ }\href {https://doi.org/10.1103/PhysRevLett.40.223} {\bibfield
  {journal} {\bibinfo  {journal} {Phys. Rev. Lett.}\ }\textbf {\bibinfo
  {volume} {40}},\ \bibinfo {pages} {223} (\bibinfo {year} {1978})}\BibitemShut
  {NoStop}%
\bibitem [{\citenamefont {Wilczek}(1978)}]{78Wilczek279-282PRL}%
  \BibitemOpen
  \bibfield  {author} {\bibinfo {author} {\bibfnamefont {F.}~\bibnamefont
  {Wilczek}},\ }\href {https://doi.org/10.1103/PhysRevLett.40.279} {\bibfield
  {journal} {\bibinfo  {journal} {Phys. Rev. Lett.}\ }\textbf {\bibinfo
  {volume} {40}},\ \bibinfo {pages} {279} (\bibinfo {year} {1978})}\BibitemShut
  {NoStop}%
\bibitem [{\citenamefont {Kim}\ and\ \citenamefont
  {Carosi}(2010)}]{Kim-2008hd}%
  \BibitemOpen
  \bibfield  {author} {\bibinfo {author} {\bibfnamefont {J.~E.}\ \bibnamefont
  {Kim}}\ and\ \bibinfo {author} {\bibfnamefont {G.}~\bibnamefont {Carosi}},\
  }\href {https://doi.org/10.1103/RevModPhys.82.557} {\bibfield  {journal}
  {\bibinfo  {journal} {Rev. Mod. Phys.}\ }\textbf {\bibinfo {volume} {82}},\
  \bibinfo {pages} {557} (\bibinfo {year} {2010})},\ \Eprint
  {https://arxiv.org/abs/0807.3125} {arXiv:0807.3125 [hep-ph]} \BibitemShut
  {NoStop}%
\bibitem [{\citenamefont {Caputo}\ and\ \citenamefont
  {Raffelt}(2024)}]{Caputo-2024oqc}%
  \BibitemOpen
  \bibfield  {author} {\bibinfo {author} {\bibfnamefont {A.}~\bibnamefont
  {Caputo}}\ and\ \bibinfo {author} {\bibfnamefont {G.}~\bibnamefont
  {Raffelt}},\ }\href {https://doi.org/10.22323/1.454.0041} {\bibfield
  {journal} {\bibinfo  {journal} {PoS}\ }\textbf {\bibinfo {volume}
  {COSMICWISPers}},\ \bibinfo {pages} {041} (\bibinfo {year} {2024})},\ \Eprint
  {https://arxiv.org/abs/2401.13728} {arXiv:2401.13728 [hep-ph]} \BibitemShut
  {NoStop}%
\bibitem [{\citenamefont {Kawasaki}\ and\ \citenamefont
  {Nakayama}(2013)}]{13Kawasaki.Nakayama69-95ARNPS}%
  \BibitemOpen
  \bibfield  {author} {\bibinfo {author} {\bibfnamefont {M.}~\bibnamefont
  {Kawasaki}}\ and\ \bibinfo {author} {\bibfnamefont {K.}~\bibnamefont
  {Nakayama}},\ }\href {https://doi.org/10.1146/annurev-nucl-102212-170536}
  {\bibfield  {journal} {\bibinfo  {journal} {Ann. Rev. Nucl. Part. Sci.}\
  }\textbf {\bibinfo {volume} {63}},\ \bibinfo {pages} {69} (\bibinfo {year}
  {2013})},\ \Eprint {https://arxiv.org/abs/1301.1123} {arXiv:1301.1123
  [hep-ph]} \BibitemShut {NoStop}%
\bibitem [{\citenamefont {Duffy}\ and\ \citenamefont {van
  Bibber}(2009)}]{Duffy-2009ig}%
  \BibitemOpen
  \bibfield  {author} {\bibinfo {author} {\bibfnamefont {L.~D.}\ \bibnamefont
  {Duffy}}\ and\ \bibinfo {author} {\bibfnamefont {K.}~\bibnamefont {van
  Bibber}},\ }\href {https://doi.org/10.1088/1367-2630/11/10/105008} {\bibfield
   {journal} {\bibinfo  {journal} {New J. Phys.}\ }\textbf {\bibinfo {volume}
  {11}},\ \bibinfo {pages} {105008} (\bibinfo {year} {2009})},\ \Eprint
  {https://arxiv.org/abs/0904.3346} {arXiv:0904.3346 [hep-ph]} \BibitemShut
  {NoStop}%
\bibitem [{\citenamefont {Marsh}(2016)}]{16Marsh1-79PR}%
  \BibitemOpen
  \bibfield  {author} {\bibinfo {author} {\bibfnamefont {D.~J.~E.}\
  \bibnamefont {Marsh}},\ }\href
  {https://doi.org/10.1016/j.physrep.2016.06.005} {\bibfield  {journal}
  {\bibinfo  {journal} {Phys. Rept.}\ }\textbf {\bibinfo {volume} {643}},\
  \bibinfo {pages} {1} (\bibinfo {year} {2016})},\ \Eprint
  {https://arxiv.org/abs/1510.07633} {arXiv:1510.07633 [astro-ph.CO]}
  \BibitemShut {NoStop}%
\bibitem [{\citenamefont {Di~Luzio}\ \emph {et~al.}(2020)\citenamefont
  {Di~Luzio}, \citenamefont {Giannotti}, \citenamefont {Nardi},\ and\
  \citenamefont {Visinelli}}]{DiLuzio-2020wdo}%
  \BibitemOpen
  \bibfield  {author} {\bibinfo {author} {\bibfnamefont {L.}~\bibnamefont
  {Di~Luzio}}, \bibinfo {author} {\bibfnamefont {M.}~\bibnamefont {Giannotti}},
  \bibinfo {author} {\bibfnamefont {E.}~\bibnamefont {Nardi}},\ and\ \bibinfo
  {author} {\bibfnamefont {L.}~\bibnamefont {Visinelli}},\ }\href
  {https://doi.org/10.1016/j.physrep.2020.06.002} {\bibfield  {journal}
  {\bibinfo  {journal} {Phys. Rept.}\ }\textbf {\bibinfo {volume} {870}},\
  \bibinfo {pages} {1} (\bibinfo {year} {2020})},\ \Eprint
  {https://arxiv.org/abs/2003.01100} {arXiv:2003.01100 [hep-ph]} \BibitemShut
  {NoStop}%
\bibitem [{\citenamefont {Di~Luzio}\ \emph {et~al.}(2022)\citenamefont
  {Di~Luzio}, \citenamefont {Fedele}, \citenamefont {Giannotti}, \citenamefont
  {Mescia},\ and\ \citenamefont {Nardi}}]{DiLuzio-2021ysg}%
  \BibitemOpen
  \bibfield  {author} {\bibinfo {author} {\bibfnamefont {L.}~\bibnamefont
  {Di~Luzio}}, \bibinfo {author} {\bibfnamefont {M.}~\bibnamefont {Fedele}},
  \bibinfo {author} {\bibfnamefont {M.}~\bibnamefont {Giannotti}}, \bibinfo
  {author} {\bibfnamefont {F.}~\bibnamefont {Mescia}},\ and\ \bibinfo {author}
  {\bibfnamefont {E.}~\bibnamefont {Nardi}},\ }\href
  {https://doi.org/10.1088/1475-7516/2022/02/035} {\bibfield  {journal}
  {\bibinfo  {journal} {JCAP}\ }\textbf {\bibinfo {volume} {02}},\ \bibinfo
  {pages} {035} (\bibinfo {year} {2022})},\ \Eprint
  {https://arxiv.org/abs/2109.10368} {arXiv:2109.10368 [hep-ph]} \BibitemShut
  {NoStop}%
\bibitem [{\citenamefont {Choi}, \citenamefont {Im},\ and\ \citenamefont
  {Shin}(2021)}]{Choi-2020rgn}%
  \BibitemOpen
  \bibfield  {author} {\bibinfo {author} {\bibfnamefont {K.}~\bibnamefont
  {Choi}}, \bibinfo {author} {\bibfnamefont {S.~H.}\ \bibnamefont {Im}},\ and\
  \bibinfo {author} {\bibfnamefont {C.~S.}\ \bibnamefont {Shin}},\ }\href
  {https://doi.org/10.1146/annurev-nucl-120720-031147} {\bibfield  {journal}
  {\bibinfo  {journal} {Annu. Rev. Nucl. Part. Sci.}\ ,\ } (\bibinfo {year}
  {2021})},\ \Eprint {https://arxiv.org/abs/2012.05029} {arXiv:2012.05029
  [hep-ph]} \BibitemShut {NoStop}%
\bibitem [{\citenamefont {Bradley}\ \emph {et~al.}(2003)\citenamefont
  {Bradley}, \citenamefont {Clarke}, \citenamefont {Kinion}, \citenamefont
  {Rosenberg}, \citenamefont {van Bibber}, \citenamefont {Matsuki},
  \citenamefont {Muck},\ and\ \citenamefont
  {Sikivie}}]{03Bradley.Clarke.ea777-817RMP}%
  \BibitemOpen
  \bibfield  {author} {\bibinfo {author} {\bibfnamefont {R.}~\bibnamefont
  {Bradley}}, \bibinfo {author} {\bibfnamefont {J.}~\bibnamefont {Clarke}},
  \bibinfo {author} {\bibfnamefont {D.}~\bibnamefont {Kinion}}, \bibinfo
  {author} {\bibfnamefont {L.~J.}\ \bibnamefont {Rosenberg}}, \bibinfo {author}
  {\bibfnamefont {K.}~\bibnamefont {van Bibber}}, \bibinfo {author}
  {\bibfnamefont {S.}~\bibnamefont {Matsuki}}, \bibinfo {author} {\bibfnamefont
  {M.}~\bibnamefont {Muck}},\ and\ \bibinfo {author} {\bibfnamefont
  {P.}~\bibnamefont {Sikivie}},\ }\href
  {https://doi.org/10.1103/RevModPhys.75.777} {\bibfield  {journal} {\bibinfo
  {journal} {Rev. Mod. Phys.}\ }\textbf {\bibinfo {volume} {75}},\ \bibinfo
  {pages} {777} (\bibinfo {year} {2003})}\BibitemShut {NoStop}%
\bibitem [{\citenamefont {Jaeckel}\ and\ \citenamefont
  {Ringwald}(2010)}]{Jaeckel-2010ni}%
  \BibitemOpen
  \bibfield  {author} {\bibinfo {author} {\bibfnamefont {J.}~\bibnamefont
  {Jaeckel}}\ and\ \bibinfo {author} {\bibfnamefont {A.}~\bibnamefont
  {Ringwald}},\ }\href {https://doi.org/10.1146/annurev.nucl.012809.104433}
  {\bibfield  {journal} {\bibinfo  {journal} {Ann. Rev. Nucl. Part. Sci.}\
  }\textbf {\bibinfo {volume} {60}},\ \bibinfo {pages} {405} (\bibinfo {year}
  {2010})},\ \Eprint {https://arxiv.org/abs/1002.0329} {arXiv:1002.0329
  [hep-ph]} \BibitemShut {NoStop}%
\bibitem [{\citenamefont {Ringwald}(2012)}]{Ringwald-2012hr}%
  \BibitemOpen
  \bibfield  {author} {\bibinfo {author} {\bibfnamefont {A.}~\bibnamefont
  {Ringwald}},\ }\href {https://doi.org/10.1016/j.dark.2012.10.008} {\bibfield
  {journal} {\bibinfo  {journal} {Phys. Dark Univ.}\ }\textbf {\bibinfo
  {volume} {1}},\ \bibinfo {pages} {116} (\bibinfo {year} {2012})},\ \Eprint
  {https://arxiv.org/abs/1210.5081} {arXiv:1210.5081 [hep-ph]} \BibitemShut
  {NoStop}%
\bibitem [{\citenamefont {Graham}\ \emph {et~al.}(2015)\citenamefont {Graham},
  \citenamefont {Irastorza}, \citenamefont {Lamoreaux}, \citenamefont
  {Lindner},\ and\ \citenamefont {van
  Bibber}}]{15Graham.Irastorza.ea485-514ARNPS}%
  \BibitemOpen
  \bibfield  {author} {\bibinfo {author} {\bibfnamefont {P.~W.}\ \bibnamefont
  {Graham}}, \bibinfo {author} {\bibfnamefont {I.~G.}\ \bibnamefont
  {Irastorza}}, \bibinfo {author} {\bibfnamefont {S.~K.}\ \bibnamefont
  {Lamoreaux}}, \bibinfo {author} {\bibfnamefont {A.}~\bibnamefont {Lindner}},\
  and\ \bibinfo {author} {\bibfnamefont {K.~A.}\ \bibnamefont {van Bibber}},\
  }\href {https://doi.org/10.1146/annurev-nucl-102014-022120} {\bibfield
  {journal} {\bibinfo  {journal} {Ann. Rev. Nucl. Part. Sci.}\ }\textbf
  {\bibinfo {volume} {65}},\ \bibinfo {pages} {485} (\bibinfo {year} {2015})},\
  \Eprint {https://arxiv.org/abs/1602.00039} {arXiv:1602.00039 [hep-ex]}
  \BibitemShut {NoStop}%
\bibitem [{\citenamefont {Irastorza}\ and\ \citenamefont
  {Redondo}(2018)}]{Irastorza-2018dyq}%
  \BibitemOpen
  \bibfield  {author} {\bibinfo {author} {\bibfnamefont {I.~G.}\ \bibnamefont
  {Irastorza}}\ and\ \bibinfo {author} {\bibfnamefont {J.}~\bibnamefont
  {Redondo}},\ }\href {https://doi.org/10.1016/j.ppnp.2018.05.003} {\bibfield
  {journal} {\bibinfo  {journal} {Prog. Part. Nucl. Phys.}\ }\textbf {\bibinfo
  {volume} {102}},\ \bibinfo {pages} {89} (\bibinfo {year} {2018})},\ \Eprint
  {https://arxiv.org/abs/1801.08127} {arXiv:1801.08127 [hep-ph]} \BibitemShut
  {NoStop}%
\bibitem [{\citenamefont {Sikivie}(2021)}]{Sikivie-2020zpn}%
  \BibitemOpen
  \bibfield  {author} {\bibinfo {author} {\bibfnamefont {P.}~\bibnamefont
  {Sikivie}},\ }\href {https://doi.org/10.1103/RevModPhys.93.015004} {\bibfield
   {journal} {\bibinfo  {journal} {Rev. Mod. Phys.}\ }\textbf {\bibinfo
  {volume} {93}},\ \bibinfo {pages} {015004} (\bibinfo {year} {2021})},\
  \Eprint {https://arxiv.org/abs/2003.02206} {arXiv:2003.02206 [hep-ph]}
  \BibitemShut {NoStop}%
\bibitem [{\citenamefont {Chen}\ and\ \citenamefont
  {Kephart}(2024)}]{Chen-2023jki}%
  \BibitemOpen
  \bibfield  {author} {\bibinfo {author} {\bibfnamefont {L.}~\bibnamefont
  {Chen}}\ and\ \bibinfo {author} {\bibfnamefont {T.~W.}\ \bibnamefont
  {Kephart}},\ }\href {https://doi.org/10.3390/universe10010024} {\bibfield
  {journal} {\bibinfo  {journal} {Universe}\ }\textbf {\bibinfo {volume}
  {10}},\ \bibinfo {pages} {24} (\bibinfo {year} {2024})},\ \Eprint
  {https://arxiv.org/abs/2311.16453} {arXiv:2311.16453 [hep-ph]} \BibitemShut
  {NoStop}%
\bibitem [{\citenamefont {Semertzidis}\ and\ \citenamefont
  {Youn}(2022)}]{Semertzidis-2021rxs}%
  \BibitemOpen
  \bibfield  {author} {\bibinfo {author} {\bibfnamefont {Y.~K.}\ \bibnamefont
  {Semertzidis}}\ and\ \bibinfo {author} {\bibfnamefont {S.}~\bibnamefont
  {Youn}},\ }\href {https://doi.org/10.1126/sciadv.abm9928} {\bibfield
  {journal} {\bibinfo  {journal} {Sci. Adv.}\ }\textbf {\bibinfo {volume}
  {8}},\ \bibinfo {pages} {abm9928} (\bibinfo {year} {2022})},\ \Eprint
  {https://arxiv.org/abs/2104.14831} {arXiv:2104.14831 [hep-ph]} \BibitemShut
  {NoStop}%
\bibitem [{\citenamefont {Antel}\ \emph {et~al.}(2023)\citenamefont {Antel}
  \emph {et~al.}}]{Antel-2023hkf}%
  \BibitemOpen
  \bibfield  {author} {\bibinfo {author} {\bibfnamefont {C.}~\bibnamefont
  {Antel}} \emph {et~al.},\ }\href
  {https://doi.org/10.1140/epjc/s10052-023-12168-5} {\bibfield  {journal}
  {\bibinfo  {journal} {Eur. Phys. J. C}\ }\textbf {\bibinfo {volume} {83}},\
  \bibinfo {pages} {1122} (\bibinfo {year} {2023})},\ \Eprint
  {https://arxiv.org/abs/2305.01715} {arXiv:2305.01715 [hep-ph]} \BibitemShut
  {NoStop}%
\bibitem [{\citenamefont {Preskill}, \citenamefont {Wise},\ and\ \citenamefont
  {Wilczek}(1983)}]{83Preskill.Wise.ea127-132PLB}%
  \BibitemOpen
  \bibfield  {author} {\bibinfo {author} {\bibfnamefont {J.}~\bibnamefont
  {Preskill}}, \bibinfo {author} {\bibfnamefont {M.~B.}\ \bibnamefont {Wise}},\
  and\ \bibinfo {author} {\bibfnamefont {F.}~\bibnamefont {Wilczek}},\ }\href
  {https://doi.org/10.1016/0370-2693(83)90637-8} {\bibfield  {journal}
  {\bibinfo  {journal} {Phys. Lett. B}\ }\textbf {\bibinfo {volume} {120}},\
  \bibinfo {pages} {127} (\bibinfo {year} {1983})}\BibitemShut {NoStop}%
\bibitem [{\citenamefont {Abbott}\ and\ \citenamefont
  {Sikivie}(1983)}]{Abbott-1982af}%
  \BibitemOpen
  \bibfield  {author} {\bibinfo {author} {\bibfnamefont {L.~F.}\ \bibnamefont
  {Abbott}}\ and\ \bibinfo {author} {\bibfnamefont {P.}~\bibnamefont
  {Sikivie}},\ }\href {https://doi.org/10.1016/0370-2693(83)90638-X} {\bibfield
   {journal} {\bibinfo  {journal} {Phys. Lett. B}\ }\textbf {\bibinfo {volume}
  {120}},\ \bibinfo {pages} {133} (\bibinfo {year} {1983})}\BibitemShut
  {NoStop}%
\bibitem [{\citenamefont {Dine}\ and\ \citenamefont
  {Fischler}(1983)}]{83DinePLB}%
  \BibitemOpen
  \bibfield  {author} {\bibinfo {author} {\bibfnamefont {M.}~\bibnamefont
  {Dine}}\ and\ \bibinfo {author} {\bibfnamefont {W.}~\bibnamefont
  {Fischler}},\ }\href
  {https://doi.org/https://doi.org/10.1016/0370-2693(83)90639-1} {\bibfield
  {journal} {\bibinfo  {journal} {Phys. Lett. B}\ }\textbf {\bibinfo {volume}
  {120}},\ \bibinfo {pages} {137 } (\bibinfo {year} {1983})}\BibitemShut
  {NoStop}%
\bibitem [{\citenamefont {O'Hare}(2024)}]{OHare-2024nmr}%
  \BibitemOpen
  \bibfield  {author} {\bibinfo {author} {\bibfnamefont {C.~A.~J.}\
  \bibnamefont {O'Hare}},\ }\href {https://doi.org/10.22323/1.454.0040}
  {\bibfield  {journal} {\bibinfo  {journal} {PoS}\ }\textbf {\bibinfo {volume}
  {COSMICWISPers}},\ \bibinfo {pages} {040} (\bibinfo {year} {2024})},\ \Eprint
  {https://arxiv.org/abs/2403.17697} {arXiv:2403.17697 [hep-ph]} \BibitemShut
  {NoStop}%
\bibitem [{\citenamefont {Balkin}\ \emph {et~al.}(2020)\citenamefont {Balkin},
  \citenamefont {Serra}, \citenamefont {Springmann},\ and\ \citenamefont
  {Weiler}}]{Balkin-2020dsr}%
  \BibitemOpen
  \bibfield  {author} {\bibinfo {author} {\bibfnamefont {R.}~\bibnamefont
  {Balkin}}, \bibinfo {author} {\bibfnamefont {J.}~\bibnamefont {Serra}},
  \bibinfo {author} {\bibfnamefont {K.}~\bibnamefont {Springmann}},\ and\
  \bibinfo {author} {\bibfnamefont {A.}~\bibnamefont {Weiler}},\ }\href
  {https://doi.org/10.1007/JHEP07(2020)221} {\bibfield  {journal} {\bibinfo
  {journal} {JHEP}\ }\textbf {\bibinfo {volume} {07}},\ \bibinfo {pages} {221}
  (\bibinfo {year} {2020})},\ \Eprint {https://arxiv.org/abs/2003.04903}
  {arXiv:2003.04903 [hep-ph]} \BibitemShut {NoStop}%
\bibitem [{\citenamefont {Anzuini}\ \emph {et~al.}(2024)\citenamefont
  {Anzuini}, \citenamefont {Gomez-Banon}, \citenamefont {Pons}, \citenamefont
  {Melatos},\ and\ \citenamefont {Lasky}}]{Anzuini-2023whm}%
  \BibitemOpen
  \bibfield  {author} {\bibinfo {author} {\bibfnamefont {F.}~\bibnamefont
  {Anzuini}}, \bibinfo {author} {\bibfnamefont {A.}~\bibnamefont
  {Gomez-Banon}}, \bibinfo {author} {\bibfnamefont {J.~A.}\ \bibnamefont
  {Pons}}, \bibinfo {author} {\bibfnamefont {A.}~\bibnamefont {Melatos}},\ and\
  \bibinfo {author} {\bibfnamefont {P.~D.}\ \bibnamefont {Lasky}},\ }\href
  {https://doi.org/10.1103/PhysRevD.109.083030} {\bibfield  {journal} {\bibinfo
   {journal} {Phys. Rev. D}\ }\textbf {\bibinfo {volume} {109}},\ \bibinfo
  {pages} {083030} (\bibinfo {year} {2024})},\ \Eprint
  {https://arxiv.org/abs/2311.13776} {arXiv:2311.13776 [hep-ph]} \BibitemShut
  {NoStop}%
\bibitem [{\citenamefont {Anzuini}\ \emph {et~al.}(2023)\citenamefont
  {Anzuini}, \citenamefont {Pons}, \citenamefont {G\'omez-Ba\~n\'on},
  \citenamefont {Lasky}, \citenamefont {Bianchini},\ and\ \citenamefont
  {Melatos}}]{Anzuini-2022bqd}%
  \BibitemOpen
  \bibfield  {author} {\bibinfo {author} {\bibfnamefont {F.}~\bibnamefont
  {Anzuini}}, \bibinfo {author} {\bibfnamefont {J.~A.}\ \bibnamefont {Pons}},
  \bibinfo {author} {\bibfnamefont {A.}~\bibnamefont {G\'omez-Ba\~n\'on}},
  \bibinfo {author} {\bibfnamefont {P.~D.}\ \bibnamefont {Lasky}}, \bibinfo
  {author} {\bibfnamefont {F.}~\bibnamefont {Bianchini}},\ and\ \bibinfo
  {author} {\bibfnamefont {A.}~\bibnamefont {Melatos}},\ }\href
  {https://doi.org/10.1103/PhysRevLett.130.071001} {\bibfield  {journal}
  {\bibinfo  {journal} {Phys. Rev. Lett.}\ }\textbf {\bibinfo {volume} {130}},\
  \bibinfo {pages} {071001} (\bibinfo {year} {2023})},\ \Eprint
  {https://arxiv.org/abs/2211.10863} {arXiv:2211.10863 [astro-ph.HE]}
  \BibitemShut {NoStop}%
\bibitem [{\citenamefont {Ferrer}\ and\ \citenamefont {de~la
  Incera}(2024)}]{Ferrer-2024xwu}%
  \BibitemOpen
  \bibfield  {author} {\bibinfo {author} {\bibfnamefont {E.~J.}\ \bibnamefont
  {Ferrer}}\ and\ \bibinfo {author} {\bibfnamefont {V.}~\bibnamefont {de~la
  Incera}},\ }\href {https://doi.org/10.1140/epjc/s10052-024-12486-2}
  {\bibfield  {journal} {\bibinfo  {journal} {Eur. Phys. J. C}\ }\textbf
  {\bibinfo {volume} {84}},\ \bibinfo {pages} {133} (\bibinfo {year} {2024})},\
  \Eprint {https://arxiv.org/abs/2403.17035} {arXiv:2403.17035 [nucl-th]}
  \BibitemShut {NoStop}%
\bibitem [{\citenamefont {Hook}\ and\ \citenamefont
  {Huang}(2018)}]{Hook-2017psm}%
  \BibitemOpen
  \bibfield  {author} {\bibinfo {author} {\bibfnamefont {A.}~\bibnamefont
  {Hook}}\ and\ \bibinfo {author} {\bibfnamefont {J.}~\bibnamefont {Huang}},\
  }\href {https://doi.org/10.1007/JHEP06(2018)036} {\bibfield  {journal}
  {\bibinfo  {journal} {JHEP}\ }\textbf {\bibinfo {volume} {06}},\ \bibinfo
  {pages} {036} (\bibinfo {year} {2018})},\ \Eprint
  {https://arxiv.org/abs/1708.08464} {arXiv:1708.08464 [hep-ph]} \BibitemShut
  {NoStop}%
\bibitem [{\citenamefont {Lopes}\ \emph {et~al.}(2022)\citenamefont {Lopes},
  \citenamefont {Farias}, \citenamefont {Dexheimer}, \citenamefont
  {Bandyopadhyay},\ and\ \citenamefont {O.~Ramos}}]{Lopes-2022efy}%
  \BibitemOpen
  \bibfield  {author} {\bibinfo {author} {\bibfnamefont {B.~S.}\ \bibnamefont
  {Lopes}}, \bibinfo {author} {\bibfnamefont {R.~L.~S.}\ \bibnamefont
  {Farias}}, \bibinfo {author} {\bibfnamefont {V.}~\bibnamefont {Dexheimer}},
  \bibinfo {author} {\bibfnamefont {A.}~\bibnamefont {Bandyopadhyay}},\ and\
  \bibinfo {author} {\bibfnamefont {R.}~\bibnamefont {O.~Ramos}},\ }\href
  {https://doi.org/10.1103/PhysRevD.106.L121301} {\bibfield  {journal}
  {\bibinfo  {journal} {Phys. Rev. D}\ }\textbf {\bibinfo {volume} {106}},\
  \bibinfo {pages} {L121301} (\bibinfo {year} {2022})},\ \Eprint
  {https://arxiv.org/abs/2206.01631} {arXiv:2206.01631 [hep-ph]} \BibitemShut
  {NoStop}%
\bibitem [{\citenamefont {Yadav}, \citenamefont {Mishra},\ and\ \citenamefont
  {Sarkar}(2024)}]{Yadav-2024xob}%
  \BibitemOpen
  \bibfield  {author} {\bibinfo {author} {\bibfnamefont {S.}~\bibnamefont
  {Yadav}}, \bibinfo {author} {\bibfnamefont {M.}~\bibnamefont {Mishra}},\ and\
  \bibinfo {author} {\bibfnamefont {T.~G.}\ \bibnamefont {Sarkar}},\ }\href
  {https://doi.org/10.1140/epjc/s10052-024-13051-7} {\bibfield  {journal}
  {\bibinfo  {journal} {Eur. Phys. J. C}\ }\textbf {\bibinfo {volume} {84}},\
  \bibinfo {pages} {687} (\bibinfo {year} {2024})},\ \Eprint
  {https://arxiv.org/abs/2403.15305} {arXiv:2403.15305 [astro-ph.HE]}
  \BibitemShut {NoStop}%
\bibitem [{\citenamefont {Cavan-Piton}\ \emph {et~al.}(2024)\citenamefont
  {Cavan-Piton}, \citenamefont {Guadagnoli}, \citenamefont {Oertel},
  \citenamefont {Seong},\ and\ \citenamefont {Vittorio}}]{Cavan-Piton-2024ayu}%
  \BibitemOpen
  \bibfield  {author} {\bibinfo {author} {\bibfnamefont {M.}~\bibnamefont
  {Cavan-Piton}}, \bibinfo {author} {\bibfnamefont {D.}~\bibnamefont
  {Guadagnoli}}, \bibinfo {author} {\bibfnamefont {M.}~\bibnamefont {Oertel}},
  \bibinfo {author} {\bibfnamefont {H.}~\bibnamefont {Seong}},\ and\ \bibinfo
  {author} {\bibfnamefont {L.}~\bibnamefont {Vittorio}},\ }\href
  {https://doi.org/10.1103/PhysRevLett.133.121002} {\bibfield  {journal}
  {\bibinfo  {journal} {Phys. Rev. Lett.}\ }\textbf {\bibinfo {volume} {133}},\
  \bibinfo {pages} {121002} (\bibinfo {year} {2024})},\ \Eprint
  {https://arxiv.org/abs/2401.10979} {arXiv:2401.10979 [hep-ph]} \BibitemShut
  {NoStop}%
\bibitem [{\citenamefont {Song}, \citenamefont {Su},\ and\ \citenamefont
  {Wu}(2024)}]{Song-2024rru}%
  \BibitemOpen
  \bibfield  {author} {\bibinfo {author} {\bibfnamefont {N.}~\bibnamefont
  {Song}}, \bibinfo {author} {\bibfnamefont {L.}~\bibnamefont {Su}},\ and\
  \bibinfo {author} {\bibfnamefont {L.}~\bibnamefont {Wu}},\ }\href@noop {} {\
  (\bibinfo {year} {2024})},\ \Eprint {https://arxiv.org/abs/2402.15144}
  {arXiv:2402.15144 [hep-ph]} \BibitemShut {NoStop}%
\bibitem [{\citenamefont {Murgana}\ \emph {et~al.}(2024)\citenamefont
  {Murgana}, \citenamefont {Castillo}, \citenamefont {Grunfeld},\ and\
  \citenamefont {Ruggieri}}]{Murgana-2024djt}%
  \BibitemOpen
  \bibfield  {author} {\bibinfo {author} {\bibfnamefont {F.}~\bibnamefont
  {Murgana}}, \bibinfo {author} {\bibfnamefont {D.~E.~A.}\ \bibnamefont
  {Castillo}}, \bibinfo {author} {\bibfnamefont {A.~G.}\ \bibnamefont
  {Grunfeld}},\ and\ \bibinfo {author} {\bibfnamefont {M.}~\bibnamefont
  {Ruggieri}},\ }\href {https://doi.org/10.1103/PhysRevD.110.014042} {\bibfield
   {journal} {\bibinfo  {journal} {Phys. Rev. D}\ }\textbf {\bibinfo {volume}
  {110}},\ \bibinfo {pages} {014042} (\bibinfo {year} {2024})},\ \Eprint
  {https://arxiv.org/abs/2404.14160} {arXiv:2404.14160 [hep-ph]} \BibitemShut
  {NoStop}%
\bibitem [{\citenamefont {Salvio}, \citenamefont {Strumia},\ and\ \citenamefont
  {Xue}(2014)}]{Salvio-2013iaa}%
  \BibitemOpen
  \bibfield  {author} {\bibinfo {author} {\bibfnamefont {A.}~\bibnamefont
  {Salvio}}, \bibinfo {author} {\bibfnamefont {A.}~\bibnamefont {Strumia}},\
  and\ \bibinfo {author} {\bibfnamefont {W.}~\bibnamefont {Xue}},\ }\href
  {https://doi.org/10.1088/1475-7516/2014/01/011} {\bibfield  {journal}
  {\bibinfo  {journal} {JCAP}\ }\textbf {\bibinfo {volume} {01}},\ \bibinfo
  {pages} {011} (\bibinfo {year} {2014})},\ \Eprint
  {https://arxiv.org/abs/1310.6982} {arXiv:1310.6982 [hep-ph]} \BibitemShut
  {NoStop}%
\bibitem [{\citenamefont {D'Eramo}, \citenamefont {Hajkarim},\ and\
  \citenamefont {Yun}(2021)}]{DEramo-2021lgb}%
  \BibitemOpen
  \bibfield  {author} {\bibinfo {author} {\bibfnamefont {F.}~\bibnamefont
  {D'Eramo}}, \bibinfo {author} {\bibfnamefont {F.}~\bibnamefont {Hajkarim}},\
  and\ \bibinfo {author} {\bibfnamefont {S.}~\bibnamefont {Yun}},\ }\href
  {https://doi.org/10.1007/JHEP10(2021)224} {\bibfield  {journal} {\bibinfo
  {journal} {JHEP}\ }\textbf {\bibinfo {volume} {10}},\ \bibinfo {pages} {224}
  (\bibinfo {year} {2021})},\ \Eprint {https://arxiv.org/abs/2108.05371}
  {arXiv:2108.05371 [hep-ph]} \BibitemShut {NoStop}%
\bibitem [{\citenamefont {Ferreira}\ and\ \citenamefont
  {Notari}(2018)}]{18Ferreira.Notari191301-191301PRL}%
  \BibitemOpen
  \bibfield  {author} {\bibinfo {author} {\bibfnamefont {R.~Z.}\ \bibnamefont
  {Ferreira}}\ and\ \bibinfo {author} {\bibfnamefont {A.}~\bibnamefont
  {Notari}},\ }\href {https://doi.org/10.1103/PhysRevLett.120.191301}
  {\bibfield  {journal} {\bibinfo  {journal} {Phys. Rev. Lett.}\ }\textbf
  {\bibinfo {volume} {120}},\ \bibinfo {pages} {191301} (\bibinfo {year}
  {2018})},\ \Eprint {https://arxiv.org/abs/1801.06090} {arXiv:1801.06090
  [hep-ph]} \BibitemShut {NoStop}%
\bibitem [{\citenamefont {Bandyopadhyay}\ \emph {et~al.}(2019)\citenamefont
  {Bandyopadhyay}, \citenamefont {Farias}, \citenamefont {Lopes},\ and\
  \citenamefont {Ramos}}]{Bandyopadhyay-2019pml}%
  \BibitemOpen
  \bibfield  {author} {\bibinfo {author} {\bibfnamefont {A.}~\bibnamefont
  {Bandyopadhyay}}, \bibinfo {author} {\bibfnamefont {R.~L.~S.}\ \bibnamefont
  {Farias}}, \bibinfo {author} {\bibfnamefont {B.~S.}\ \bibnamefont {Lopes}},\
  and\ \bibinfo {author} {\bibfnamefont {R.~O.}\ \bibnamefont {Ramos}},\ }\href
  {https://doi.org/10.1103/PhysRevD.100.076021} {\bibfield  {journal} {\bibinfo
   {journal} {Phys. Rev. D}\ }\textbf {\bibinfo {volume} {100}},\ \bibinfo
  {pages} {076021} (\bibinfo {year} {2019})},\ \Eprint
  {https://arxiv.org/abs/1906.09250} {arXiv:1906.09250 [hep-ph]} \BibitemShut
  {NoStop}%
\bibitem [{\citenamefont {Notari}, \citenamefont {Rompineve},\ and\
  \citenamefont {Villadoro}(2023)}]{Notari-2022ffe}%
  \BibitemOpen
  \bibfield  {author} {\bibinfo {author} {\bibfnamefont {A.}~\bibnamefont
  {Notari}}, \bibinfo {author} {\bibfnamefont {F.}~\bibnamefont {Rompineve}},\
  and\ \bibinfo {author} {\bibfnamefont {G.}~\bibnamefont {Villadoro}},\ }\href
  {https://doi.org/10.1103/PhysRevLett.131.011004} {\bibfield  {journal}
  {\bibinfo  {journal} {Phys. Rev. Lett.}\ }\textbf {\bibinfo {volume} {131}},\
  \bibinfo {pages} {011004} (\bibinfo {year} {2023})},\ \Eprint
  {https://arxiv.org/abs/2211.03799} {arXiv:2211.03799 [hep-ph]} \BibitemShut
  {NoStop}%
\bibitem [{\citenamefont {D'Eramo}, \citenamefont {Hajkarim},\ and\
  \citenamefont {Yun}(2022)}]{DEramo-2021psx}%
  \BibitemOpen
  \bibfield  {author} {\bibinfo {author} {\bibfnamefont {F.}~\bibnamefont
  {D'Eramo}}, \bibinfo {author} {\bibfnamefont {F.}~\bibnamefont {Hajkarim}},\
  and\ \bibinfo {author} {\bibfnamefont {S.}~\bibnamefont {Yun}},\ }\href
  {https://doi.org/10.1103/PhysRevLett.128.152001} {\bibfield  {journal}
  {\bibinfo  {journal} {Phys. Rev. Lett.}\ }\textbf {\bibinfo {volume} {128}},\
  \bibinfo {pages} {152001} (\bibinfo {year} {2022})},\ \Eprint
  {https://arxiv.org/abs/2108.04259} {arXiv:2108.04259 [hep-ph]} \BibitemShut
  {NoStop}%
\bibitem [{\citenamefont {Ferreira}, \citenamefont {Notari},\ and\
  \citenamefont {Rompineve}(2021)}]{Ferreira-2020bpb}%
  \BibitemOpen
  \bibfield  {author} {\bibinfo {author} {\bibfnamefont {R.~Z.}\ \bibnamefont
  {Ferreira}}, \bibinfo {author} {\bibfnamefont {A.}~\bibnamefont {Notari}},\
  and\ \bibinfo {author} {\bibfnamefont {F.}~\bibnamefont {Rompineve}},\ }\href
  {https://doi.org/10.1103/PhysRevD.103.063524} {\bibfield  {journal} {\bibinfo
   {journal} {Phys. Rev. D}\ }\textbf {\bibinfo {volume} {103}},\ \bibinfo
  {pages} {063524} (\bibinfo {year} {2021})},\ \Eprint
  {https://arxiv.org/abs/2012.06566} {arXiv:2012.06566 [hep-ph]} \BibitemShut
  {NoStop}%
\bibitem [{\citenamefont {D'Eramo}\ \emph {et~al.}(2022)\citenamefont
  {D'Eramo}, \citenamefont {Di~Valentino}, \citenamefont {Giar\`e},
  \citenamefont {Hajkarim}, \citenamefont {Melchiorri}, \citenamefont {Mena},
  \citenamefont {Renzi},\ and\ \citenamefont {Yun}}]{DEramo-2022nvb}%
  \BibitemOpen
  \bibfield  {author} {\bibinfo {author} {\bibfnamefont {F.}~\bibnamefont
  {D'Eramo}}, \bibinfo {author} {\bibfnamefont {E.}~\bibnamefont
  {Di~Valentino}}, \bibinfo {author} {\bibfnamefont {W.}~\bibnamefont
  {Giar\`e}}, \bibinfo {author} {\bibfnamefont {F.}~\bibnamefont {Hajkarim}},
  \bibinfo {author} {\bibfnamefont {A.}~\bibnamefont {Melchiorri}}, \bibinfo
  {author} {\bibfnamefont {O.}~\bibnamefont {Mena}}, \bibinfo {author}
  {\bibfnamefont {F.}~\bibnamefont {Renzi}},\ and\ \bibinfo {author}
  {\bibfnamefont {S.}~\bibnamefont {Yun}},\ }\href
  {https://doi.org/10.1088/1475-7516/2022/09/022} {\bibfield  {journal}
  {\bibinfo  {journal} {JCAP}\ }\textbf {\bibinfo {volume} {09}},\ \bibinfo
  {pages} {022} (\bibinfo {year} {2022})},\ \Eprint
  {https://arxiv.org/abs/2205.07849} {arXiv:2205.07849 [astro-ph.CO]}
  \BibitemShut {NoStop}%
\bibitem [{\citenamefont {D'Eramo}\ \emph {et~al.}(2018)\citenamefont
  {D'Eramo}, \citenamefont {Ferreira}, \citenamefont {Notari},\ and\
  \citenamefont {Bernal}}]{DEramo-2018vss}%
  \BibitemOpen
  \bibfield  {author} {\bibinfo {author} {\bibfnamefont {F.}~\bibnamefont
  {D'Eramo}}, \bibinfo {author} {\bibfnamefont {R.~Z.}\ \bibnamefont
  {Ferreira}}, \bibinfo {author} {\bibfnamefont {A.}~\bibnamefont {Notari}},\
  and\ \bibinfo {author} {\bibfnamefont {J.~L.}\ \bibnamefont {Bernal}},\
  }\href {https://doi.org/10.1088/1475-7516/2018/11/014} {\bibfield  {journal}
  {\bibinfo  {journal} {JCAP}\ }\textbf {\bibinfo {volume} {1811}},\ \bibinfo
  {pages} {014} (\bibinfo {year} {2018})},\ \Eprint
  {https://arxiv.org/abs/1808.07430} {arXiv:1808.07430 [hep-ph]} \BibitemShut
  {NoStop}%
\bibitem [{\citenamefont {Badziak}\ \emph {et~al.}(2024)\citenamefont
  {Badziak}, \citenamefont {Harigaya}, \citenamefont {\L{}ukawski},\ and\
  \citenamefont {Ziegler}}]{Badziak-2024szg}%
  \BibitemOpen
  \bibfield  {author} {\bibinfo {author} {\bibfnamefont {M.}~\bibnamefont
  {Badziak}}, \bibinfo {author} {\bibfnamefont {K.}~\bibnamefont {Harigaya}},
  \bibinfo {author} {\bibfnamefont {M.}~\bibnamefont {\L{}ukawski}},\ and\
  \bibinfo {author} {\bibfnamefont {R.}~\bibnamefont {Ziegler}},\ }\href@noop
  {} {\  (\bibinfo {year} {2024})},\ \Eprint {https://arxiv.org/abs/2403.05621}
  {arXiv:2403.05621 [hep-ph]} \BibitemShut {NoStop}%
\bibitem [{\citenamefont {Wang}, \citenamefont {Guo},\ and\ \citenamefont
  {Zhou}(2024)}]{Wang-2023xny}%
  \BibitemOpen
  \bibfield  {author} {\bibinfo {author} {\bibfnamefont {J.-B.}\ \bibnamefont
  {Wang}}, \bibinfo {author} {\bibfnamefont {Z.-H.}\ \bibnamefont {Guo}},\ and\
  \bibinfo {author} {\bibfnamefont {H.-Q.}\ \bibnamefont {Zhou}},\ }\href
  {https://doi.org/10.1103/PhysRevD.109.075030} {\bibfield  {journal} {\bibinfo
   {journal} {Phys. Rev. D}\ }\textbf {\bibinfo {volume} {109}},\ \bibinfo
  {pages} {075030} (\bibinfo {year} {2024})},\ \Eprint
  {https://arxiv.org/abs/2312.15240} {arXiv:2312.15240 [hep-ph]} \BibitemShut
  {NoStop}%
\bibitem [{\citenamefont {Splittorff}\ and\ \citenamefont
  {Verbaarschot}(2007)}]{Splittorff-2007ck}%
  \BibitemOpen
  \bibfield  {author} {\bibinfo {author} {\bibfnamefont {K.}~\bibnamefont
  {Splittorff}}\ and\ \bibinfo {author} {\bibfnamefont {J.~J.~M.}\ \bibnamefont
  {Verbaarschot}},\ }\href {https://doi.org/10.1103/PhysRevD.75.116003}
  {\bibfield  {journal} {\bibinfo  {journal} {Phys. Rev. D}\ }\textbf {\bibinfo
  {volume} {75}},\ \bibinfo {pages} {116003} (\bibinfo {year} {2007})},\
  \Eprint {https://arxiv.org/abs/hep-lat/0702011} {arXiv:hep-lat/0702011}
  \BibitemShut {NoStop}%
\bibitem [{\citenamefont {Prosperi}, \citenamefont {Raciti},\ and\
  \citenamefont {Simolo}(2007)}]{Prosperi-2006hx}%
  \BibitemOpen
  \bibfield  {author} {\bibinfo {author} {\bibfnamefont {G.}~\bibnamefont
  {Prosperi}}, \bibinfo {author} {\bibfnamefont {M.}~\bibnamefont {Raciti}},\
  and\ \bibinfo {author} {\bibfnamefont {C.}~\bibnamefont {Simolo}},\ }\href
  {https://doi.org/10.1016/j.ppnp.2006.09.001} {\bibfield  {journal} {\bibinfo
  {journal} {Prog. Part. Nucl. Phys.}\ }\textbf {\bibinfo {volume} {58}},\
  \bibinfo {pages} {387} (\bibinfo {year} {2007})},\ \Eprint
  {https://arxiv.org/abs/hep-ph/0607209} {arXiv:hep-ph/0607209} \BibitemShut
  {NoStop}%
\bibitem [{\citenamefont {Deur}, \citenamefont {Brodsky},\ and\ \citenamefont
  {de~Teramond}(2016)}]{Deur-2016tte}%
  \BibitemOpen
  \bibfield  {author} {\bibinfo {author} {\bibfnamefont {A.}~\bibnamefont
  {Deur}}, \bibinfo {author} {\bibfnamefont {S.~J.}\ \bibnamefont {Brodsky}},\
  and\ \bibinfo {author} {\bibfnamefont {G.~F.}\ \bibnamefont {de~Teramond}},\
  }\href {https://doi.org/10.1016/j.ppnp.2016.04.003} {\bibfield  {journal}
  {\bibinfo  {journal} {Prog. Part. Nucl. Phys.}\ }\textbf {\bibinfo {volume}
  {90}},\ \bibinfo {pages} {1} (\bibinfo {year} {2016})},\ \Eprint
  {https://arxiv.org/abs/1604.08082} {arXiv:1604.08082 [hep-ph]} \BibitemShut
  {NoStop}%
\bibitem [{\citenamefont {Nambu}\ and\ \citenamefont
  {Jona-Lasinio}(1961{\natexlab{a}})}]{Nambu-1961fr}%
  \BibitemOpen
  \bibfield  {author} {\bibinfo {author} {\bibfnamefont {Y.}~\bibnamefont
  {Nambu}}\ and\ \bibinfo {author} {\bibfnamefont {G.}~\bibnamefont
  {Jona-Lasinio}},\ }\href {https://doi.org/10.1103/PhysRev.124.246} {\bibfield
   {journal} {\bibinfo  {journal} {Phys. Rev.}\ }\textbf {\bibinfo {volume}
  {124}},\ \bibinfo {pages} {246} (\bibinfo {year}
  {1961}{\natexlab{a}})}\BibitemShut {NoStop}%
\bibitem [{\citenamefont {Nambu}\ and\ \citenamefont
  {Jona-Lasinio}(1961{\natexlab{b}})}]{61Nambu.Jona-Lasinio345-358PR}%
  \BibitemOpen
  \bibfield  {author} {\bibinfo {author} {\bibfnamefont {Y.}~\bibnamefont
  {Nambu}}\ and\ \bibinfo {author} {\bibfnamefont {G.}~\bibnamefont
  {Jona-Lasinio}},\ }\href {https://doi.org/10.1103/PhysRev.122.345} {\bibfield
   {journal} {\bibinfo  {journal} {Phys. Rev.}\ }\textbf {\bibinfo {volume}
  {122}},\ \bibinfo {pages} {345} (\bibinfo {year}
  {1961}{\natexlab{b}})}\BibitemShut {NoStop}%
\bibitem [{\citenamefont {Hatsuda}\ and\ \citenamefont
  {Kunihiro}(1994)}]{94Hatsuda.Kunihiro221-367PR}%
  \BibitemOpen
  \bibfield  {author} {\bibinfo {author} {\bibfnamefont {T.}~\bibnamefont
  {Hatsuda}}\ and\ \bibinfo {author} {\bibfnamefont {T.}~\bibnamefont
  {Kunihiro}},\ }\href {https://doi.org/10.1016/0370-1573(94)90022-1}
  {\bibfield  {journal} {\bibinfo  {journal} {Phys. Rept.}\ }\textbf {\bibinfo
  {volume} {247}},\ \bibinfo {pages} {221} (\bibinfo {year} {1994})},\ \Eprint
  {https://arxiv.org/abs/hep-ph/9401310} {arXiv:hep-ph/9401310 [hep-ph]}
  \BibitemShut {NoStop}%
\bibitem [{\citenamefont {Klevansky}(1992)}]{92Klevansky649-708RMP}%
  \BibitemOpen
  \bibfield  {author} {\bibinfo {author} {\bibfnamefont {S.~P.}\ \bibnamefont
  {Klevansky}},\ }\href {https://doi.org/10.1103/RevModPhys.64.649} {\bibfield
  {journal} {\bibinfo  {journal} {Rev. Mod. Phys.}\ }\textbf {\bibinfo {volume}
  {64}},\ \bibinfo {pages} {649} (\bibinfo {year} {1992})}\BibitemShut
  {NoStop}%
\bibitem [{\citenamefont {Buballa}(2005)}]{05Buballa-PhysRep}%
  \BibitemOpen
  \bibfield  {author} {\bibinfo {author} {\bibfnamefont {M.}~\bibnamefont
  {Buballa}},\ }\href
  {https://doi.org/https://doi.org/10.1016/j.physrep.2004.11.004} {\bibfield
  {journal} {\bibinfo  {journal} {Phys. Rep.}\ }\textbf {\bibinfo {volume}
  {407}},\ \bibinfo {pages} {205 } (\bibinfo {year} {2005})}\BibitemShut
  {NoStop}%
\bibitem [{\citenamefont {Volkov}\ and\ \citenamefont
  {Radzhabov}(2006)}]{Volkov-2005kw}%
  \BibitemOpen
  \bibfield  {author} {\bibinfo {author} {\bibfnamefont {M.~K.}\ \bibnamefont
  {Volkov}}\ and\ \bibinfo {author} {\bibfnamefont {A.~E.}\ \bibnamefont
  {Radzhabov}},\ }\href {https://doi.org/10.1070/PU2006v049n06ABEH005905}
  {\bibfield  {journal} {\bibinfo  {journal} {Phys. Usp.}\ }\textbf {\bibinfo
  {volume} {49}},\ \bibinfo {pages} {551} (\bibinfo {year} {2006})},\ \Eprint
  {https://arxiv.org/abs/hep-ph/0508263} {arXiv:hep-ph/0508263} \BibitemShut
  {NoStop}%
\bibitem [{\citenamefont {Lu}, \citenamefont {Xia},\ and\ \citenamefont
  {Ruggieri}(2020)}]{Lu-2019diy}%
  \BibitemOpen
  \bibfield  {author} {\bibinfo {author} {\bibfnamefont {Z.-Y.}\ \bibnamefont
  {Lu}}, \bibinfo {author} {\bibfnamefont {C.-J.}\ \bibnamefont {Xia}},\ and\
  \bibinfo {author} {\bibfnamefont {M.}~\bibnamefont {Ruggieri}},\ }\href
  {https://doi.org/10.1140/epjc/s10052-020-7614-6} {\bibfield  {journal}
  {\bibinfo  {journal} {Eur. Phys. J. C}\ }\textbf {\bibinfo {volume} {80}},\
  \bibinfo {pages} {46} (\bibinfo {year} {2020})},\ \Eprint
  {https://arxiv.org/abs/1907.11497} {arXiv:1907.11497 [hep-ph]} \BibitemShut
  {NoStop}%
\bibitem [{\citenamefont {Lu}\ \emph {et~al.}(2021)\citenamefont {Lu},
  \citenamefont {Chen}, \citenamefont {Lu}, \citenamefont {Xu},\ and\
  \citenamefont {Li}}]{Lu-2021hvw}%
  \BibitemOpen
  \bibfield  {author} {\bibinfo {author} {\bibfnamefont {Q.}~\bibnamefont
  {Lu}}, \bibinfo {author} {\bibfnamefont {W.-J.}\ \bibnamefont {Chen}},
  \bibinfo {author} {\bibfnamefont {Z.-Y.}\ \bibnamefont {Lu}}, \bibinfo
  {author} {\bibfnamefont {Y.}~\bibnamefont {Xu}},\ and\ \bibinfo {author}
  {\bibfnamefont {X.-Q.}\ \bibnamefont {Li}},\ }\href
  {https://doi.org/10.7498/aps.70.20210132} {\bibfield  {journal} {\bibinfo
  {journal} {Acta Phys. Sin.}\ }\textbf {\bibinfo {volume} {70}},\ \bibinfo
  {pages} {145101} (\bibinfo {year} {2021})}\BibitemShut {NoStop}%
\bibitem [{\citenamefont {Carignano}, \citenamefont {Mammarella},\ and\
  \citenamefont {Mannarelli}(2016)}]{16Carignano.Mammarella.ea51503-51503PRD}%
  \BibitemOpen
  \bibfield  {author} {\bibinfo {author} {\bibfnamefont {S.}~\bibnamefont
  {Carignano}}, \bibinfo {author} {\bibfnamefont {A.}~\bibnamefont
  {Mammarella}},\ and\ \bibinfo {author} {\bibfnamefont {M.}~\bibnamefont
  {Mannarelli}},\ }\href {https://doi.org/10.1103/PhysRevD.93.051503}
  {\bibfield  {journal} {\bibinfo  {journal} {Phys. Rev. D}\ }\textbf {\bibinfo
  {volume} {93}},\ \bibinfo {pages} {051503} (\bibinfo {year} {2016})},\
  \Eprint {https://arxiv.org/abs/1602.01317} {arXiv:1602.01317 [hep-ph]}
  \BibitemShut {NoStop}%
\bibitem [{\citenamefont {Detmold}, \citenamefont {Orginos},\ and\
  \citenamefont {Shi}(2012)}]{12Detmold.Orginos.ea54507-54507PRD}%
  \BibitemOpen
  \bibfield  {author} {\bibinfo {author} {\bibfnamefont {W.}~\bibnamefont
  {Detmold}}, \bibinfo {author} {\bibfnamefont {K.}~\bibnamefont {Orginos}},\
  and\ \bibinfo {author} {\bibfnamefont {Z.}~\bibnamefont {Shi}},\ }\href
  {https://doi.org/10.1103/PhysRevD.86.054507} {\bibfield  {journal} {\bibinfo
  {journal} {Phys. Rev. D}\ }\textbf {\bibinfo {volume} {86}},\ \bibinfo
  {pages} {054507} (\bibinfo {year} {2012})},\ \Eprint
  {https://arxiv.org/abs/1205.4224} {arXiv:1205.4224 [hep-lat]} \BibitemShut
  {NoStop}%
\bibitem [{\citenamefont {He}, \citenamefont {Jin},\ and\ \citenamefont
  {Zhuang}(2005)}]{05He.Jin.ea116001-116001PRD}%
  \BibitemOpen
  \bibfield  {author} {\bibinfo {author} {\bibfnamefont {L.-y.}\ \bibnamefont
  {He}}, \bibinfo {author} {\bibfnamefont {M.}~\bibnamefont {Jin}},\ and\
  \bibinfo {author} {\bibfnamefont {P.-f.}\ \bibnamefont {Zhuang}},\ }\href
  {https://doi.org/10.1103/PhysRevD.71.116001} {\bibfield  {journal} {\bibinfo
  {journal} {Phys. Rev. D}\ }\textbf {\bibinfo {volume} {71}},\ \bibinfo
  {pages} {116001} (\bibinfo {year} {2005})},\ \Eprint
  {https://arxiv.org/abs/hep-ph/0503272} {arXiv:hep-ph/0503272 [hep-ph]}
  \BibitemShut {NoStop}%
\bibitem [{\citenamefont {Kogut}\ and\ \citenamefont
  {Sinclair}(2002)}]{02Kogut.Sinclair34505-34505PRD}%
  \BibitemOpen
  \bibfield  {author} {\bibinfo {author} {\bibfnamefont {J.~B.}\ \bibnamefont
  {Kogut}}\ and\ \bibinfo {author} {\bibfnamefont {D.~K.}\ \bibnamefont
  {Sinclair}},\ }\href {https://doi.org/10.1103/PhysRevD.66.034505} {\bibfield
  {journal} {\bibinfo  {journal} {Phys. Rev. D}\ }\textbf {\bibinfo {volume}
  {66}},\ \bibinfo {pages} {034505} (\bibinfo {year} {2002})},\ \Eprint
  {https://arxiv.org/abs/hep-lat/0202028} {arXiv:hep-lat/0202028 [hep-lat]}
  \BibitemShut {NoStop}%
\bibitem [{\citenamefont {Kogut}\ and\ \citenamefont
  {Sinclair}(2004)}]{04Kogut.Sinclair94501-94501PRD}%
  \BibitemOpen
  \bibfield  {author} {\bibinfo {author} {\bibfnamefont {J.~B.}\ \bibnamefont
  {Kogut}}\ and\ \bibinfo {author} {\bibfnamefont {D.~K.}\ \bibnamefont
  {Sinclair}},\ }\href {https://doi.org/10.1103/PhysRevD.70.094501} {\bibfield
  {journal} {\bibinfo  {journal} {Phys. Rev. D}\ }\textbf {\bibinfo {volume}
  {70}},\ \bibinfo {pages} {094501} (\bibinfo {year} {2004})},\ \Eprint
  {https://arxiv.org/abs/hep-lat/0407027} {arXiv:hep-lat/0407027 [hep-lat]}
  \BibitemShut {NoStop}%
\bibitem [{\citenamefont {Son}\ and\ \citenamefont
  {Stephanov}(2001)}]{01Son.Stephanov592-595PRL}%
  \BibitemOpen
  \bibfield  {author} {\bibinfo {author} {\bibfnamefont {D.~T.}\ \bibnamefont
  {Son}}\ and\ \bibinfo {author} {\bibfnamefont {M.~A.}\ \bibnamefont
  {Stephanov}},\ }\href {https://doi.org/10.1103/PhysRevLett.86.592} {\bibfield
   {journal} {\bibinfo  {journal} {Phys. Rev. Lett.}\ }\textbf {\bibinfo
  {volume} {86}},\ \bibinfo {pages} {592} (\bibinfo {year} {2001})},\ \Eprint
  {https://arxiv.org/abs/hep-ph/0005225} {arXiv:hep-ph/0005225 [hep-ph]}
  \BibitemShut {NoStop}%
\bibitem [{\citenamefont {Lu}\ and\ \citenamefont
  {Ruggieri}(2019)}]{Lu-2018ukl}%
  \BibitemOpen
  \bibfield  {author} {\bibinfo {author} {\bibfnamefont {Z.-Y.}\ \bibnamefont
  {Lu}}\ and\ \bibinfo {author} {\bibfnamefont {M.}~\bibnamefont {Ruggieri}},\
  }\href {https://doi.org/10.1103/PhysRevD.100.014013} {\bibfield  {journal}
  {\bibinfo  {journal} {Phys. Rev. D}\ }\textbf {\bibinfo {volume} {100}},\
  \bibinfo {pages} {014013} (\bibinfo {year} {2019})},\ \Eprint
  {https://arxiv.org/abs/1811.05102} {arXiv:1811.05102 [hep-ph]} \BibitemShut
  {NoStop}%
\bibitem [{\citenamefont {Grilli~di Cortona}\ \emph {et~al.}(2016)\citenamefont
  {Grilli~di Cortona}, \citenamefont {Hardy}, \citenamefont {Pardo~Vega},\ and\
  \citenamefont {Villadoro}}]{GrillidiCortona-2015jxo}%
  \BibitemOpen
  \bibfield  {author} {\bibinfo {author} {\bibfnamefont {G.}~\bibnamefont
  {Grilli~di Cortona}}, \bibinfo {author} {\bibfnamefont {E.}~\bibnamefont
  {Hardy}}, \bibinfo {author} {\bibfnamefont {J.}~\bibnamefont {Pardo~Vega}},\
  and\ \bibinfo {author} {\bibfnamefont {G.}~\bibnamefont {Villadoro}},\ }\href
  {https://doi.org/10.1007/JHEP01(2016)034} {\bibfield  {journal} {\bibinfo
  {journal} {JHEP}\ }\textbf {\bibinfo {volume} {01}},\ \bibinfo {pages} {034}
  (\bibinfo {year} {2016})},\ \Eprint {https://arxiv.org/abs/1511.02867}
  {arXiv:1511.02867 [hep-ph]} \BibitemShut {NoStop}%
\bibitem [{\citenamefont {Borsanyi}\ \emph
  {et~al.}(2016{\natexlab{a}})\citenamefont {Borsanyi} \emph
  {et~al.}}]{16Borsanyi.others69-71N}%
  \BibitemOpen
  \bibfield  {author} {\bibinfo {author} {\bibfnamefont {S.}~\bibnamefont
  {Borsanyi}} \emph {et~al.},\ }\href {https://doi.org/10.1038/nature20115}
  {\bibfield  {journal} {\bibinfo  {journal} {Nature}\ }\textbf {\bibinfo
  {volume} {539}},\ \bibinfo {pages} {69} (\bibinfo {year}
  {2016}{\natexlab{a}})},\ \Eprint {https://arxiv.org/abs/1606.07494}
  {arXiv:1606.07494 [hep-lat]} \BibitemShut {NoStop}%
\bibitem [{\citenamefont {Boomsma}\ and\ \citenamefont
  {Boer}(2009)}]{09Boomsma.Boer34019-34019PRD}%
  \BibitemOpen
  \bibfield  {author} {\bibinfo {author} {\bibfnamefont {J.~K.}\ \bibnamefont
  {Boomsma}}\ and\ \bibinfo {author} {\bibfnamefont {D.}~\bibnamefont {Boer}},\
  }\href {https://doi.org/10.1103/PhysRevD.80.034019} {\bibfield  {journal}
  {\bibinfo  {journal} {Phys. Rev. D}\ }\textbf {\bibinfo {volume} {80}},\
  \bibinfo {pages} {034019} (\bibinfo {year} {2009})},\ \Eprint
  {https://arxiv.org/abs/0905.4660} {arXiv:0905.4660 [hep-ph]} \BibitemShut
  {NoStop}%
\bibitem [{\citenamefont {Das}, \citenamefont {Mishra},\ and\ \citenamefont
  {Mohapatra}(2021)}]{Das-2020pjg}%
  \BibitemOpen
  \bibfield  {author} {\bibinfo {author} {\bibfnamefont {A.}~\bibnamefont
  {Das}}, \bibinfo {author} {\bibfnamefont {H.}~\bibnamefont {Mishra}},\ and\
  \bibinfo {author} {\bibfnamefont {R.~K.}\ \bibnamefont {Mohapatra}},\ }\href
  {https://doi.org/10.1103/PhysRevD.103.074003} {\bibfield  {journal} {\bibinfo
   {journal} {Phys. Rev. D}\ }\textbf {\bibinfo {volume} {103}},\ \bibinfo
  {pages} {074003} (\bibinfo {year} {2021})},\ \Eprint
  {https://arxiv.org/abs/2006.15727} {arXiv:2006.15727 [hep-ph]} \BibitemShut
  {NoStop}%
\bibitem [{\citenamefont {Xu}\ \emph {et~al.}(2015)\citenamefont {Xu},
  \citenamefont {Peng}, \citenamefont {Lu},\ and\ \citenamefont
  {Cui}}]{Xu-2014zea}%
  \BibitemOpen
  \bibfield  {author} {\bibinfo {author} {\bibfnamefont {J.-F.}\ \bibnamefont
  {Xu}}, \bibinfo {author} {\bibfnamefont {G.-X.}\ \bibnamefont {Peng}},
  \bibinfo {author} {\bibfnamefont {Z.-Y.}\ \bibnamefont {Lu}},\ and\ \bibinfo
  {author} {\bibfnamefont {S.-S.}\ \bibnamefont {Cui}},\ }\href
  {https://doi.org/10.1007/s11433-014-5599-6} {\bibfield  {journal} {\bibinfo
  {journal} {Sci. China Phys. Mech. Astron.}\ }\textbf {\bibinfo {volume}
  {58}},\ \bibinfo {pages} {042001} (\bibinfo {year} {2015})}\BibitemShut
  {NoStop}%
\bibitem [{\citenamefont {Ma}\ \emph {et~al.}(2023)\citenamefont {Ma},
  \citenamefont {Lu}, \citenamefont {Xu}, \citenamefont {Peng}, \citenamefont
  {Fu},\ and\ \citenamefont {Wang}}]{Ma-2023stj}%
  \BibitemOpen
  \bibfield  {author} {\bibinfo {author} {\bibfnamefont {Z.-J.}\ \bibnamefont
  {Ma}}, \bibinfo {author} {\bibfnamefont {Z.-Y.}\ \bibnamefont {Lu}}, \bibinfo
  {author} {\bibfnamefont {J.-F.}\ \bibnamefont {Xu}}, \bibinfo {author}
  {\bibfnamefont {G.-X.}\ \bibnamefont {Peng}}, \bibinfo {author}
  {\bibfnamefont {X.}~\bibnamefont {Fu}},\ and\ \bibinfo {author}
  {\bibfnamefont {J.}~\bibnamefont {Wang}},\ }\href
  {https://doi.org/10.1103/PhysRevD.108.054017} {\bibfield  {journal} {\bibinfo
   {journal} {Phys. Rev. D}\ }\textbf {\bibinfo {volume} {108}},\ \bibinfo
  {pages} {054017} (\bibinfo {year} {2023})},\ \Eprint
  {https://arxiv.org/abs/2308.05308} {arXiv:2308.05308 [hep-ph]} \BibitemShut
  {NoStop}%
\bibitem [{\citenamefont {Karsch}(2002)}]{Karsch-2001cy}%
  \BibitemOpen
  \bibfield  {author} {\bibinfo {author} {\bibfnamefont {F.}~\bibnamefont
  {Karsch}},\ }\href {https://doi.org/10.1007/3-540-45792-5_6} {\bibfield
  {journal} {\bibinfo  {journal} {Lect. Notes Phys.}\ }\textbf {\bibinfo
  {volume} {583}},\ \bibinfo {pages} {209} (\bibinfo {year} {2002})},\ \Eprint
  {https://arxiv.org/abs/hep-lat/0106019} {arXiv:hep-lat/0106019} \BibitemShut
  {NoStop}%
\bibitem [{\citenamefont {Fodor}\ and\ \citenamefont
  {Katz}(2002)}]{Fodor-2001ye}%
  \BibitemOpen
  \bibfield  {author} {\bibinfo {author} {\bibfnamefont {Z.}~\bibnamefont
  {Fodor}}\ and\ \bibinfo {author} {\bibfnamefont {S.~D.}\ \bibnamefont {Katz}}
  (\bibinfo {collaboration} {C01-08-19}),\ }\href
  {https://doi.org/10.1016/S0920-5632(01)01740-6} {\bibfield  {journal}
  {\bibinfo  {journal} {Nucl. Phys. B Proc. Suppl.}\ }\textbf {\bibinfo
  {volume} {106}},\ \bibinfo {pages} {441} (\bibinfo {year} {2002})},\ \Eprint
  {https://arxiv.org/abs/hep-lat/0110102} {arXiv:hep-lat/0110102} \BibitemShut
  {NoStop}%
\bibitem [{\citenamefont {Schaefer}\ and\ \citenamefont
  {Wagner}(2009)}]{Schaefer-2008hk}%
  \BibitemOpen
  \bibfield  {author} {\bibinfo {author} {\bibfnamefont {B.-J.}\ \bibnamefont
  {Schaefer}}\ and\ \bibinfo {author} {\bibfnamefont {M.}~\bibnamefont
  {Wagner}},\ }\href {https://doi.org/10.1103/PhysRevD.79.014018} {\bibfield
  {journal} {\bibinfo  {journal} {Phys. Rev. D}\ }\textbf {\bibinfo {volume}
  {79}},\ \bibinfo {pages} {014018} (\bibinfo {year} {2009})},\ \Eprint
  {https://arxiv.org/abs/0808.1491} {arXiv:0808.1491 [hep-ph]} \BibitemShut
  {NoStop}%
\bibitem [{\citenamefont {Fukushima}, \citenamefont {Ohnishi},\ and\
  \citenamefont {Ohta}(2001)}]{Fukushima-2001hr}%
  \BibitemOpen
  \bibfield  {author} {\bibinfo {author} {\bibfnamefont {K.}~\bibnamefont
  {Fukushima}}, \bibinfo {author} {\bibfnamefont {K.}~\bibnamefont {Ohnishi}},\
  and\ \bibinfo {author} {\bibfnamefont {K.}~\bibnamefont {Ohta}},\ }\href
  {https://doi.org/10.1103/PhysRevC.63.045203} {\bibfield  {journal} {\bibinfo
  {journal} {Phys. Rev. C}\ }\textbf {\bibinfo {volume} {63}},\ \bibinfo
  {pages} {045203} (\bibinfo {year} {2001})},\ \Eprint
  {https://arxiv.org/abs/nucl-th/0101062} {arXiv:nucl-th/0101062 [nucl-th]}
  \BibitemShut {NoStop}%
\bibitem [{\citenamefont {Jiang}, \citenamefont {Xia},\ and\ \citenamefont
  {Zhuang}(2016)}]{16Jiang.Xia.ea74006-74006PRD}%
  \BibitemOpen
  \bibfield  {author} {\bibinfo {author} {\bibfnamefont {Y.}~\bibnamefont
  {Jiang}}, \bibinfo {author} {\bibfnamefont {T.}~\bibnamefont {Xia}},\ and\
  \bibinfo {author} {\bibfnamefont {P.}~\bibnamefont {Zhuang}},\ }\href
  {https://doi.org/10.1103/PhysRevD.93.074006} {\bibfield  {journal} {\bibinfo
  {journal} {Phys. Rev. D}\ }\textbf {\bibinfo {volume} {93}},\ \bibinfo
  {pages} {074006} (\bibinfo {year} {2016})},\ \Eprint
  {https://arxiv.org/abs/1511.06466} {arXiv:1511.06466 [hep-ph]} \BibitemShut
  {NoStop}%
\bibitem [{\citenamefont {Iida}, \citenamefont {Itou},\ and\ \citenamefont
  {Lee}(2020)}]{Iida-2019rah}%
  \BibitemOpen
  \bibfield  {author} {\bibinfo {author} {\bibfnamefont {K.}~\bibnamefont
  {Iida}}, \bibinfo {author} {\bibfnamefont {E.}~\bibnamefont {Itou}},\ and\
  \bibinfo {author} {\bibfnamefont {T.-G.}\ \bibnamefont {Lee}},\ }\href
  {https://doi.org/10.1007/JHEP01(2020)181} {\bibfield  {journal} {\bibinfo
  {journal} {JHEP}\ }\textbf {\bibinfo {volume} {01}},\ \bibinfo {pages} {181}
  (\bibinfo {year} {2020})},\ \Eprint {https://arxiv.org/abs/1910.07872}
  {arXiv:1910.07872 [hep-lat]} \BibitemShut {NoStop}%
\bibitem [{\citenamefont {Vicari}\ and\ \citenamefont
  {Panagopoulos}(2009)}]{Vicari-2008jw}%
  \BibitemOpen
  \bibfield  {author} {\bibinfo {author} {\bibfnamefont {E.}~\bibnamefont
  {Vicari}}\ and\ \bibinfo {author} {\bibfnamefont {H.}~\bibnamefont
  {Panagopoulos}},\ }\href {https://doi.org/10.1016/j.physrep.2008.10.001}
  {\bibfield  {journal} {\bibinfo  {journal} {Phys. Rept.}\ }\textbf {\bibinfo
  {volume} {470}},\ \bibinfo {pages} {93} (\bibinfo {year} {2009})},\ \Eprint
  {https://arxiv.org/abs/0803.1593} {arXiv:0803.1593 [hep-th]} \BibitemShut
  {NoStop}%
\bibitem [{\citenamefont {Bonati}, \citenamefont {D'Elia},\ and\ \citenamefont
  {Scapellato}(2016)}]{16Bonati.DElia.ea25028-25028PRD}%
  \BibitemOpen
  \bibfield  {author} {\bibinfo {author} {\bibfnamefont {C.}~\bibnamefont
  {Bonati}}, \bibinfo {author} {\bibfnamefont {M.}~\bibnamefont {D'Elia}},\
  and\ \bibinfo {author} {\bibfnamefont {A.}~\bibnamefont {Scapellato}},\
  }\href {https://doi.org/10.1103/PhysRevD.93.025028} {\bibfield  {journal}
  {\bibinfo  {journal} {Phys. Rev. D}\ }\textbf {\bibinfo {volume} {93}},\
  \bibinfo {pages} {025028} (\bibinfo {year} {2016})},\ \Eprint
  {https://arxiv.org/abs/1512.01544} {arXiv:1512.01544 [hep-lat]} \BibitemShut
  {NoStop}%
\bibitem [{\citenamefont {Bonati}\ \emph {et~al.}(2016)\citenamefont {Bonati},
  \citenamefont {D'Elia}, \citenamefont {Mariti}, \citenamefont {Martinelli},
  \citenamefont {Mesiti}, \citenamefont {Negro}, \citenamefont {Sanfilippo},\
  and\ \citenamefont {Villadoro}}]{Bonati-2015vqz}%
  \BibitemOpen
  \bibfield  {author} {\bibinfo {author} {\bibfnamefont {C.}~\bibnamefont
  {Bonati}}, \bibinfo {author} {\bibfnamefont {M.}~\bibnamefont {D'Elia}},
  \bibinfo {author} {\bibfnamefont {M.}~\bibnamefont {Mariti}}, \bibinfo
  {author} {\bibfnamefont {G.}~\bibnamefont {Martinelli}}, \bibinfo {author}
  {\bibfnamefont {M.}~\bibnamefont {Mesiti}}, \bibinfo {author} {\bibfnamefont
  {F.}~\bibnamefont {Negro}}, \bibinfo {author} {\bibfnamefont
  {F.}~\bibnamefont {Sanfilippo}},\ and\ \bibinfo {author} {\bibfnamefont
  {G.}~\bibnamefont {Villadoro}},\ }\href
  {https://doi.org/10.1007/JHEP03(2016)155} {\bibfield  {journal} {\bibinfo
  {journal} {JHEP}\ }\textbf {\bibinfo {volume} {03}},\ \bibinfo {pages} {155}
  (\bibinfo {year} {2016})},\ \Eprint {https://arxiv.org/abs/1512.06746}
  {arXiv:1512.06746 [hep-lat]} \BibitemShut {NoStop}%
\bibitem [{\citenamefont {Bonati}\ \emph {et~al.}(2013)\citenamefont {Bonati},
  \citenamefont {D'Elia}, \citenamefont {Panagopoulos},\ and\ \citenamefont
  {Vicari}}]{Bonati-2013tt}%
  \BibitemOpen
  \bibfield  {author} {\bibinfo {author} {\bibfnamefont {C.}~\bibnamefont
  {Bonati}}, \bibinfo {author} {\bibfnamefont {M.}~\bibnamefont {D'Elia}},
  \bibinfo {author} {\bibfnamefont {H.}~\bibnamefont {Panagopoulos}},\ and\
  \bibinfo {author} {\bibfnamefont {E.}~\bibnamefont {Vicari}},\ }\href
  {https://doi.org/10.1103/PhysRevLett.110.252003} {\bibfield  {journal}
  {\bibinfo  {journal} {Phys. Rev. Lett.}\ }\textbf {\bibinfo {volume} {110}},\
  \bibinfo {pages} {252003} (\bibinfo {year} {2013})},\ \Eprint
  {https://arxiv.org/abs/1301.7640} {arXiv:1301.7640 [hep-lat]} \BibitemShut
  {NoStop}%
\bibitem [{\citenamefont {Bonati}(2015)}]{15Bonati6-6J}%
  \BibitemOpen
  \bibfield  {author} {\bibinfo {author} {\bibfnamefont {C.}~\bibnamefont
  {Bonati}},\ }\href {https://doi.org/10.1007/JHEP03(2015)006} {\bibfield
  {journal} {\bibinfo  {journal} {JHEP}\ }\textbf {\bibinfo {volume} {03}},\
  \bibinfo {pages} {006} (\bibinfo {year} {2015})},\ \Eprint
  {https://arxiv.org/abs/1501.01172} {arXiv:1501.01172 [hep-lat]} \BibitemShut
  {NoStop}%
\bibitem [{\citenamefont {Borsanyi}\ \emph
  {et~al.}(2016{\natexlab{b}})\citenamefont {Borsanyi}, \citenamefont
  {Dierigl}, \citenamefont {Fodor}, \citenamefont {Katz}, \citenamefont
  {Mages}, \citenamefont {Nogradi}, \citenamefont {Redondo}, \citenamefont
  {Ringwald},\ and\ \citenamefont {Szabo}}]{Borsanyi-2015cka}%
  \BibitemOpen
  \bibfield  {author} {\bibinfo {author} {\bibfnamefont {S.}~\bibnamefont
  {Borsanyi}}, \bibinfo {author} {\bibfnamefont {M.}~\bibnamefont {Dierigl}},
  \bibinfo {author} {\bibfnamefont {Z.}~\bibnamefont {Fodor}}, \bibinfo
  {author} {\bibfnamefont {S.~D.}\ \bibnamefont {Katz}}, \bibinfo {author}
  {\bibfnamefont {S.~W.}\ \bibnamefont {Mages}}, \bibinfo {author}
  {\bibfnamefont {D.}~\bibnamefont {Nogradi}}, \bibinfo {author} {\bibfnamefont
  {J.}~\bibnamefont {Redondo}}, \bibinfo {author} {\bibfnamefont
  {A.}~\bibnamefont {Ringwald}},\ and\ \bibinfo {author} {\bibfnamefont
  {K.~K.}\ \bibnamefont {Szabo}},\ }\href
  {https://doi.org/10.1016/j.physletb.2015.11.020} {\bibfield  {journal}
  {\bibinfo  {journal} {Phys. Lett. B}\ }\textbf {\bibinfo {volume} {752}},\
  \bibinfo {pages} {175} (\bibinfo {year} {2016}{\natexlab{b}})},\ \Eprint
  {https://arxiv.org/abs/1508.06917} {arXiv:1508.06917 [hep-lat]} \BibitemShut
  {NoStop}%
\bibitem [{\citenamefont {Berkowitz}, \citenamefont {Buchoff},\ and\
  \citenamefont {Rinaldi}(2015)}]{Berkowitz-2015aua}%
  \BibitemOpen
  \bibfield  {author} {\bibinfo {author} {\bibfnamefont {E.}~\bibnamefont
  {Berkowitz}}, \bibinfo {author} {\bibfnamefont {M.~I.}\ \bibnamefont
  {Buchoff}},\ and\ \bibinfo {author} {\bibfnamefont {E.}~\bibnamefont
  {Rinaldi}},\ }\href {https://doi.org/10.1103/PhysRevD.92.034507} {\bibfield
  {journal} {\bibinfo  {journal} {Phys. Rev. D}\ }\textbf {\bibinfo {volume}
  {92}},\ \bibinfo {pages} {034507} (\bibinfo {year} {2015})},\ \Eprint
  {https://arxiv.org/abs/1505.07455} {arXiv:1505.07455 [hep-ph]} \BibitemShut
  {NoStop}%
\bibitem [{\citenamefont {Petreczky}, \citenamefont {Schadler},\ and\
  \citenamefont {Sharma}(2016)}]{Petreczky-2016vrs}%
  \BibitemOpen
  \bibfield  {author} {\bibinfo {author} {\bibfnamefont {P.}~\bibnamefont
  {Petreczky}}, \bibinfo {author} {\bibfnamefont {H.-P.}\ \bibnamefont
  {Schadler}},\ and\ \bibinfo {author} {\bibfnamefont {S.}~\bibnamefont
  {Sharma}},\ }\href {https://doi.org/10.1016/j.physletb.2016.09.063}
  {\bibfield  {journal} {\bibinfo  {journal} {Phys. Lett. B}\ }\textbf
  {\bibinfo {volume} {762}},\ \bibinfo {pages} {498} (\bibinfo {year}
  {2016})},\ \Eprint {https://arxiv.org/abs/1606.03145} {arXiv:1606.03145
  [hep-lat]} \BibitemShut {NoStop}%
\bibitem [{\citenamefont {Bonati}\ \emph {et~al.}(2018)\citenamefont {Bonati},
  \citenamefont {D'Elia}, \citenamefont {Martinelli}, \citenamefont {Negro},
  \citenamefont {Sanfilippo},\ and\ \citenamefont {Todaro}}]{Bonati-2018blm}%
  \BibitemOpen
  \bibfield  {author} {\bibinfo {author} {\bibfnamefont {C.}~\bibnamefont
  {Bonati}}, \bibinfo {author} {\bibfnamefont {M.}~\bibnamefont {D'Elia}},
  \bibinfo {author} {\bibfnamefont {G.}~\bibnamefont {Martinelli}}, \bibinfo
  {author} {\bibfnamefont {F.}~\bibnamefont {Negro}}, \bibinfo {author}
  {\bibfnamefont {F.}~\bibnamefont {Sanfilippo}},\ and\ \bibinfo {author}
  {\bibfnamefont {A.}~\bibnamefont {Todaro}},\ }\href
  {https://doi.org/10.1007/JHEP11(2018)170} {\bibfield  {journal} {\bibinfo
  {journal} {JHEP}\ }\textbf {\bibinfo {volume} {11}},\ \bibinfo {pages} {170}
  (\bibinfo {year} {2018})},\ \Eprint {https://arxiv.org/abs/1807.07954}
  {arXiv:1807.07954 [hep-lat]} \BibitemShut {NoStop}%
\bibitem [{\citenamefont {Gorghetto}\ and\ \citenamefont
  {Villadoro}(2019)}]{Gorghetto-2018ocs}%
  \BibitemOpen
  \bibfield  {author} {\bibinfo {author} {\bibfnamefont {M.}~\bibnamefont
  {Gorghetto}}\ and\ \bibinfo {author} {\bibfnamefont {G.}~\bibnamefont
  {Villadoro}},\ }\href {https://doi.org/10.1007/JHEP03(2019)033} {\bibfield
  {journal} {\bibinfo  {journal} {JHEP}\ }\textbf {\bibinfo {volume} {03}},\
  \bibinfo {pages} {033} (\bibinfo {year} {2019})},\ \Eprint
  {https://arxiv.org/abs/1812.01008} {arXiv:1812.01008 [hep-ph]} \BibitemShut
  {NoStop}%
\bibitem [{\citenamefont {Horvati\'c}, \citenamefont {Kekez},\ and\
  \citenamefont {Klabucar}(2019)}]{Horvatic-2019eok}%
  \BibitemOpen
  \bibfield  {author} {\bibinfo {author} {\bibfnamefont {D.}~\bibnamefont
  {Horvati\'c}}, \bibinfo {author} {\bibfnamefont {D.}~\bibnamefont {Kekez}},\
  and\ \bibinfo {author} {\bibfnamefont {D.}~\bibnamefont {Klabucar}},\ }\href
  {https://doi.org/10.3390/universe5100208} {\bibfield  {journal} {\bibinfo
  {journal} {Universe}\ }\textbf {\bibinfo {volume} {5}},\ \bibinfo {pages}
  {208} (\bibinfo {year} {2019})}\BibitemShut {NoStop}%
\bibitem [{\citenamefont {Lu}\ \emph {et~al.}(2020)\citenamefont {Lu},
  \citenamefont {Du}, \citenamefont {Guo}, \citenamefont {Mei\ss{}ner},\ and\
  \citenamefont {Vonk}}]{Lu-2020rhp}%
  \BibitemOpen
  \bibfield  {author} {\bibinfo {author} {\bibfnamefont {Z.-Y.}\ \bibnamefont
  {Lu}}, \bibinfo {author} {\bibfnamefont {M.-L.}\ \bibnamefont {Du}}, \bibinfo
  {author} {\bibfnamefont {F.-K.}\ \bibnamefont {Guo}}, \bibinfo {author}
  {\bibfnamefont {U.-G.}\ \bibnamefont {Mei\ss{}ner}},\ and\ \bibinfo {author}
  {\bibfnamefont {T.}~\bibnamefont {Vonk}},\ }\href
  {https://doi.org/10.1007/JHEP05(2020)001} {\bibfield  {journal} {\bibinfo
  {journal} {JHEP}\ }\textbf {\bibinfo {volume} {05}},\ \bibinfo {pages} {001}
  (\bibinfo {year} {2020})},\ \Eprint {https://arxiv.org/abs/2003.01625}
  {arXiv:2003.01625 [hep-ph]} \BibitemShut {NoStop}%
\bibitem [{\citenamefont {Kim}(1987)}]{87Kim1-177PR}%
  \BibitemOpen
  \bibfield  {author} {\bibinfo {author} {\bibfnamefont {J.~E.}\ \bibnamefont
  {Kim}},\ }\href {https://doi.org/10.1016/0370-1573(87)90017-2} {\bibfield
  {journal} {\bibinfo  {journal} {Phys. Rept.}\ }\textbf {\bibinfo {volume}
  {150}},\ \bibinfo {pages} {1} (\bibinfo {year} {1987})}\BibitemShut {NoStop}%
\bibitem [{\citenamefont {Cheng}(1988)}]{Cheng-1987gp}%
  \BibitemOpen
  \bibfield  {author} {\bibinfo {author} {\bibfnamefont {H.-Y.}\ \bibnamefont
  {Cheng}},\ }\href {https://doi.org/10.1016/0370-1573(88)90135-4} {\bibfield
  {journal} {\bibinfo  {journal} {Phys. Rept.}\ }\textbf {\bibinfo {volume}
  {158}},\ \bibinfo {pages} {1} (\bibinfo {year} {1988})}\BibitemShut {NoStop}%
\bibitem [{\citenamefont {Turner}(1990)}]{90Turner67-97PR}%
  \BibitemOpen
  \bibfield  {author} {\bibinfo {author} {\bibfnamefont {M.~S.}\ \bibnamefont
  {Turner}},\ }\bibfield  {booktitle} {\emph {\bibinfo {booktitle} {{BNL
  Workshop: Axions 1989:0001-38}}},\ }\href
  {https://doi.org/10.1016/0370-1573(90)90172-X} {\bibfield  {journal}
  {\bibinfo  {journal} {Phys. Rept.}\ }\textbf {\bibinfo {volume} {197}},\
  \bibinfo {pages} {67} (\bibinfo {year} {1990})}\BibitemShut {NoStop}%
\bibitem [{\citenamefont {Raffelt}(1990)}]{90Raffelt1-113PR}%
  \BibitemOpen
  \bibfield  {author} {\bibinfo {author} {\bibfnamefont {G.~G.}\ \bibnamefont
  {Raffelt}},\ }\href {https://doi.org/10.1016/0370-1573(90)90054-6} {\bibfield
   {journal} {\bibinfo  {journal} {Phys. Rept.}\ }\textbf {\bibinfo {volume}
  {198}},\ \bibinfo {pages} {1} (\bibinfo {year} {1990})}\BibitemShut {NoStop}%
\bibitem [{\citenamefont {Dine}, \citenamefont {Fischler},\ and\ \citenamefont
  {Srednicki}(1981)}]{81Dine.Fischler.ea199-202PLB}%
  \BibitemOpen
  \bibfield  {author} {\bibinfo {author} {\bibfnamefont {M.}~\bibnamefont
  {Dine}}, \bibinfo {author} {\bibfnamefont {W.}~\bibnamefont {Fischler}},\
  and\ \bibinfo {author} {\bibfnamefont {M.}~\bibnamefont {Srednicki}},\ }\href
  {https://doi.org/10.1016/0370-2693(81)90590-6} {\bibfield  {journal}
  {\bibinfo  {journal} {Phys. Lett. B}\ }\textbf {\bibinfo {volume} {104}},\
  \bibinfo {pages} {199} (\bibinfo {year} {1981})}\BibitemShut {NoStop}%
\bibitem [{\citenamefont {Zhitnitsky}(1980)}]{Zhitnitsky-1980tq}%
  \BibitemOpen
  \bibfield  {author} {\bibinfo {author} {\bibfnamefont {A.~R.}\ \bibnamefont
  {Zhitnitsky}},\ }\href@noop {} {\bibfield  {journal} {\bibinfo  {journal}
  {Sov. J. Nucl. Phys.}\ }\textbf {\bibinfo {volume} {31}},\ \bibinfo {pages}
  {260} (\bibinfo {year} {1980})}\BibitemShut {NoStop}%
\bibitem [{\citenamefont {Kim}(1979)}]{Kim-1979if}%
  \BibitemOpen
  \bibfield  {author} {\bibinfo {author} {\bibfnamefont {J.~E.}\ \bibnamefont
  {Kim}},\ }\href {https://doi.org/10.1103/PhysRevLett.43.103} {\bibfield
  {journal} {\bibinfo  {journal} {Phys. Rev. Lett.}\ }\textbf {\bibinfo
  {volume} {43}},\ \bibinfo {pages} {103} (\bibinfo {year} {1979})}\BibitemShut
  {NoStop}%
\bibitem [{\citenamefont {Shifman}, \citenamefont {Vainshtein},\ and\
  \citenamefont {Zakharov}(1980)}]{Shifman-1979if}%
  \BibitemOpen
  \bibfield  {author} {\bibinfo {author} {\bibfnamefont {M.~A.}\ \bibnamefont
  {Shifman}}, \bibinfo {author} {\bibfnamefont {A.~I.}\ \bibnamefont
  {Vainshtein}},\ and\ \bibinfo {author} {\bibfnamefont {V.~I.}\ \bibnamefont
  {Zakharov}},\ }\href {https://doi.org/10.1016/0550-3213(80)90209-6}
  {\bibfield  {journal} {\bibinfo  {journal} {Nucl. Phys. B}\ }\textbf
  {\bibinfo {volume} {166}},\ \bibinfo {pages} {493} (\bibinfo {year}
  {1980})}\BibitemShut {NoStop}%
\bibitem [{\citenamefont {Georgi}, \citenamefont {Hall},\ and\ \citenamefont
  {Wise}(1981)}]{Georgi-1981pu}%
  \BibitemOpen
  \bibfield  {author} {\bibinfo {author} {\bibfnamefont {H.~M.}\ \bibnamefont
  {Georgi}}, \bibinfo {author} {\bibfnamefont {L.~J.}\ \bibnamefont {Hall}},\
  and\ \bibinfo {author} {\bibfnamefont {M.~B.}\ \bibnamefont {Wise}},\ }\href
  {https://doi.org/10.1016/0550-3213(81)90433-8} {\bibfield  {journal}
  {\bibinfo  {journal} {Nucl. Phys. B}\ }\textbf {\bibinfo {volume} {192}},\
  \bibinfo {pages} {409} (\bibinfo {year} {1981})}\BibitemShut {NoStop}%
\bibitem [{\citenamefont {Kamionkowski}\ and\ \citenamefont
  {March-Russell}(1992)}]{Kamionkowski-1992mf}%
  \BibitemOpen
  \bibfield  {author} {\bibinfo {author} {\bibfnamefont {M.}~\bibnamefont
  {Kamionkowski}}\ and\ \bibinfo {author} {\bibfnamefont {J.}~\bibnamefont
  {March-Russell}},\ }\href {https://doi.org/10.1016/0370-2693(92)90492-M}
  {\bibfield  {journal} {\bibinfo  {journal} {Phys. Lett. B}\ }\textbf
  {\bibinfo {volume} {282}},\ \bibinfo {pages} {137} (\bibinfo {year}
  {1992})},\ \Eprint {https://arxiv.org/abs/hep-th/9202003}
  {arXiv:hep-th/9202003} \BibitemShut {NoStop}%
\bibitem [{\citenamefont {Holman}\ \emph {et~al.}(1992)\citenamefont {Holman},
  \citenamefont {Hsu}, \citenamefont {Kephart}, \citenamefont {Kolb},
  \citenamefont {Watkins},\ and\ \citenamefont {Widrow}}]{Holman-1992us}%
  \BibitemOpen
  \bibfield  {author} {\bibinfo {author} {\bibfnamefont {R.}~\bibnamefont
  {Holman}}, \bibinfo {author} {\bibfnamefont {S.~D.~H.}\ \bibnamefont {Hsu}},
  \bibinfo {author} {\bibfnamefont {T.~W.}\ \bibnamefont {Kephart}}, \bibinfo
  {author} {\bibfnamefont {E.~W.}\ \bibnamefont {Kolb}}, \bibinfo {author}
  {\bibfnamefont {R.}~\bibnamefont {Watkins}},\ and\ \bibinfo {author}
  {\bibfnamefont {L.~M.}\ \bibnamefont {Widrow}},\ }\href
  {https://doi.org/10.1016/0370-2693(92)90491-L} {\bibfield  {journal}
  {\bibinfo  {journal} {Phys. Lett. B}\ }\textbf {\bibinfo {volume} {282}},\
  \bibinfo {pages} {132} (\bibinfo {year} {1992})},\ \Eprint
  {https://arxiv.org/abs/hep-ph/9203206} {arXiv:hep-ph/9203206} \BibitemShut
  {NoStop}%
\bibitem [{\citenamefont {Rubakov}(1997)}]{Rubakov-1997vp}%
  \BibitemOpen
  \bibfield  {author} {\bibinfo {author} {\bibfnamefont {V.~A.}\ \bibnamefont
  {Rubakov}},\ }\href {https://doi.org/10.1134/1.567390} {\bibfield  {journal}
  {\bibinfo  {journal} {JETP Lett.}\ }\textbf {\bibinfo {volume} {65}},\
  \bibinfo {pages} {621} (\bibinfo {year} {1997})},\ \Eprint
  {https://arxiv.org/abs/hep-ph/9703409} {arXiv:hep-ph/9703409} \BibitemShut
  {NoStop}%
\bibitem [{\citenamefont {Berezhiani}, \citenamefont {Gianfagna},\ and\
  \citenamefont {Giannotti}(2001)}]{Berezhiani-2000gh}%
  \BibitemOpen
  \bibfield  {author} {\bibinfo {author} {\bibfnamefont {Z.}~\bibnamefont
  {Berezhiani}}, \bibinfo {author} {\bibfnamefont {L.}~\bibnamefont
  {Gianfagna}},\ and\ \bibinfo {author} {\bibfnamefont {M.}~\bibnamefont
  {Giannotti}},\ }\href {https://doi.org/10.1016/S0370-2693(00)01392-7}
  {\bibfield  {journal} {\bibinfo  {journal} {Phys. Lett. B}\ }\textbf
  {\bibinfo {volume} {500}},\ \bibinfo {pages} {286} (\bibinfo {year}
  {2001})},\ \Eprint {https://arxiv.org/abs/hep-ph/0009290}
  {arXiv:hep-ph/0009290} \BibitemShut {NoStop}%
\bibitem [{\citenamefont {Alves}\ and\ \citenamefont
  {Weiner}(2018)}]{Alves-2017avw}%
  \BibitemOpen
  \bibfield  {author} {\bibinfo {author} {\bibfnamefont {D.~S.~M.}\
  \bibnamefont {Alves}}\ and\ \bibinfo {author} {\bibfnamefont
  {N.}~\bibnamefont {Weiner}},\ }\href
  {https://doi.org/10.1007/JHEP07(2018)092} {\bibfield  {journal} {\bibinfo
  {journal} {JHEP}\ }\textbf {\bibinfo {volume} {07}},\ \bibinfo {pages} {092}
  (\bibinfo {year} {2018})},\ \Eprint {https://arxiv.org/abs/1710.03764}
  {arXiv:1710.03764 [hep-ph]} \BibitemShut {NoStop}%
\bibitem [{\citenamefont {Zhang}\ \emph {et~al.}(2023)\citenamefont {Zhang},
  \citenamefont {Castillo}, \citenamefont {Grunfeld},\ and\ \citenamefont
  {Ruggieri}}]{Zhang-2023lij}%
  \BibitemOpen
  \bibfield  {author} {\bibinfo {author} {\bibfnamefont {B.}~\bibnamefont
  {Zhang}}, \bibinfo {author} {\bibfnamefont {D.~E.~A.}\ \bibnamefont
  {Castillo}}, \bibinfo {author} {\bibfnamefont {A.~G.}\ \bibnamefont
  {Grunfeld}},\ and\ \bibinfo {author} {\bibfnamefont {M.}~\bibnamefont
  {Ruggieri}},\ }\href {https://doi.org/10.1103/PhysRevD.108.054010} {\bibfield
   {journal} {\bibinfo  {journal} {Phys. Rev. D}\ }\textbf {\bibinfo {volume}
  {108}},\ \bibinfo {pages} {054010} (\bibinfo {year} {2023})},\ \Eprint
  {https://arxiv.org/abs/2304.10240} {arXiv:2304.10240 [hep-ph]} \BibitemShut
  {NoStop}%
\bibitem [{\citenamefont {Sikivie}\ and\ \citenamefont
  {Yang}(2009)}]{Sikivie-2009qn}%
  \BibitemOpen
  \bibfield  {author} {\bibinfo {author} {\bibfnamefont {P.}~\bibnamefont
  {Sikivie}}\ and\ \bibinfo {author} {\bibfnamefont {Q.}~\bibnamefont {Yang}},\
  }\href {https://doi.org/10.1103/PhysRevLett.103.111301} {\bibfield  {journal}
  {\bibinfo  {journal} {Phys. Rev. Lett.}\ }\textbf {\bibinfo {volume} {103}},\
  \bibinfo {pages} {111301} (\bibinfo {year} {2009})},\ \Eprint
  {https://arxiv.org/abs/0901.1106} {arXiv:0901.1106 [hep-ph]} \BibitemShut
  {NoStop}%
\bibitem [{\citenamefont {Tkachev}(1986)}]{Tkachev-1986tr}%
  \BibitemOpen
  \bibfield  {author} {\bibinfo {author} {\bibfnamefont {I.~I.}\ \bibnamefont
  {Tkachev}},\ }\href@noop {} {\bibfield  {journal} {\bibinfo  {journal} {Sov.
  Astron. Lett.}\ }\textbf {\bibinfo {volume} {12}},\ \bibinfo {pages} {305}
  (\bibinfo {year} {1986})}\BibitemShut {NoStop}%
\bibitem [{\citenamefont {Kolb}\ and\ \citenamefont
  {Tkachev}(1993)}]{Kolb-1993zz}%
  \BibitemOpen
  \bibfield  {author} {\bibinfo {author} {\bibfnamefont {E.~W.}\ \bibnamefont
  {Kolb}}\ and\ \bibinfo {author} {\bibfnamefont {I.~I.}\ \bibnamefont
  {Tkachev}},\ }\href {https://doi.org/10.1103/PhysRevLett.71.3051} {\bibfield
  {journal} {\bibinfo  {journal} {Phys. Rev. Lett.}\ }\textbf {\bibinfo
  {volume} {71}},\ \bibinfo {pages} {3051} (\bibinfo {year} {1993})},\ \Eprint
  {https://arxiv.org/abs/hep-ph/9303313} {arXiv:hep-ph/9303313 [hep-ph]}
  \BibitemShut {NoStop}%
\bibitem [{\citenamefont {Hogan}\ and\ \citenamefont
  {Rees}(1988)}]{Hogan-1988mp}%
  \BibitemOpen
  \bibfield  {author} {\bibinfo {author} {\bibfnamefont {C.~J.}\ \bibnamefont
  {Hogan}}\ and\ \bibinfo {author} {\bibfnamefont {M.~J.}\ \bibnamefont
  {Rees}},\ }\href {https://doi.org/10.1016/0370-2693(88)91655-3} {\bibfield
  {journal} {\bibinfo  {journal} {Phys. Lett. B}\ }\textbf {\bibinfo {volume}
  {205}},\ \bibinfo {pages} {228} (\bibinfo {year} {1988})}\BibitemShut
  {NoStop}%
\bibitem [{\citenamefont {Barranco}\ and\ \citenamefont
  {Bernal}(2011)}]{11Barranco.Bernal43525-43525PRD}%
  \BibitemOpen
  \bibfield  {author} {\bibinfo {author} {\bibfnamefont {J.}~\bibnamefont
  {Barranco}}\ and\ \bibinfo {author} {\bibfnamefont {A.}~\bibnamefont
  {Bernal}},\ }\href {https://doi.org/10.1103/PhysRevD.83.043525} {\bibfield
  {journal} {\bibinfo  {journal} {Phys. Rev. D}\ }\textbf {\bibinfo {volume}
  {83}},\ \bibinfo {pages} {043525} (\bibinfo {year} {2011})},\ \Eprint
  {https://arxiv.org/abs/1001.1769} {arXiv:1001.1769 [astro-ph.CO]}
  \BibitemShut {NoStop}%
\bibitem [{\citenamefont {Chavanis}(2011)}]{11Chavanis43531-43531PRD}%
  \BibitemOpen
  \bibfield  {author} {\bibinfo {author} {\bibfnamefont {P.-H.}\ \bibnamefont
  {Chavanis}},\ }\href {https://doi.org/10.1103/PhysRevD.84.043531} {\bibfield
  {journal} {\bibinfo  {journal} {Phys. Rev. D}\ }\textbf {\bibinfo {volume}
  {84}},\ \bibinfo {pages} {043531} (\bibinfo {year} {2011})},\ \Eprint
  {https://arxiv.org/abs/1103.2050} {arXiv:1103.2050 [astro-ph.CO]}
  \BibitemShut {NoStop}%
\bibitem [{\citenamefont {Zhang}(2019)}]{Zhang-2018slz}%
  \BibitemOpen
  \bibfield  {author} {\bibinfo {author} {\bibfnamefont {H.}~\bibnamefont
  {Zhang}},\ }\href {https://doi.org/10.3390/sym12010025} {\bibfield  {journal}
  {\bibinfo  {journal} {Symmetry}\ }\textbf {\bibinfo {volume} {12}},\ \bibinfo
  {pages} {25} (\bibinfo {year} {2019})},\ \Eprint
  {https://arxiv.org/abs/1810.11473} {arXiv:1810.11473 [hep-ph]} \BibitemShut
  {NoStop}%
\bibitem [{\citenamefont {Raffelt}(2008)}]{Raffelt-2006cw}%
  \BibitemOpen
  \bibfield  {author} {\bibinfo {author} {\bibfnamefont {G.~G.}\ \bibnamefont
  {Raffelt}},\ }\href {https://doi.org/10.1007/978-3-540-73518-2_3} {\bibfield
  {journal} {\bibinfo  {journal} {Lect. Notes Phys.}\ }\textbf {\bibinfo
  {volume} {741}},\ \bibinfo {pages} {51} (\bibinfo {year} {2008})},\ \Eprint
  {https://arxiv.org/abs/hep-ph/0611350} {arXiv:hep-ph/0611350} \BibitemShut
  {NoStop}%
\bibitem [{\citenamefont {Sedrakian}(2016)}]{Sedrakian-2015krq}%
  \BibitemOpen
  \bibfield  {author} {\bibinfo {author} {\bibfnamefont {A.}~\bibnamefont
  {Sedrakian}},\ }\href {https://doi.org/10.1103/PhysRevD.93.065044} {\bibfield
   {journal} {\bibinfo  {journal} {Phys. Rev. D}\ }\textbf {\bibinfo {volume}
  {93}},\ \bibinfo {pages} {065044} (\bibinfo {year} {2016})},\ \Eprint
  {https://arxiv.org/abs/1512.07828} {arXiv:1512.07828 [astro-ph.HE]}
  \BibitemShut {NoStop}%
\bibitem [{\citenamefont {Bramante}\ and\ \citenamefont
  {Raj}(2024)}]{Bramante-2023djs}%
  \BibitemOpen
  \bibfield  {author} {\bibinfo {author} {\bibfnamefont {J.}~\bibnamefont
  {Bramante}}\ and\ \bibinfo {author} {\bibfnamefont {N.}~\bibnamefont {Raj}},\
  }\href {https://doi.org/10.1016/j.physrep.2023.12.001} {\bibfield  {journal}
  {\bibinfo  {journal} {Phys. Rept.}\ }\textbf {\bibinfo {volume} {1052}},\
  \bibinfo {pages} {1} (\bibinfo {year} {2024})},\ \Eprint
  {https://arxiv.org/abs/2307.14435} {arXiv:2307.14435 [hep-ph]} \BibitemShut
  {NoStop}%
\bibitem [{\citenamefont {Karkevandi}\ \emph {et~al.}(2024)\citenamefont
  {Karkevandi}, \citenamefont {Shahrbaf}, \citenamefont {Shakeri},\ and\
  \citenamefont {Typel}}]{Karkevandi-2024vov}%
  \BibitemOpen
  \bibfield  {author} {\bibinfo {author} {\bibfnamefont {D.~R.}\ \bibnamefont
  {Karkevandi}}, \bibinfo {author} {\bibfnamefont {M.}~\bibnamefont
  {Shahrbaf}}, \bibinfo {author} {\bibfnamefont {S.}~\bibnamefont {Shakeri}},\
  and\ \bibinfo {author} {\bibfnamefont {S.}~\bibnamefont {Typel}},\ }\href
  {https://doi.org/10.3390/particles7010011} {\bibfield  {journal} {\bibinfo
  {journal} {Particles}\ }\textbf {\bibinfo {volume} {7}},\ \bibinfo {pages}
  {201} (\bibinfo {year} {2024})},\ \Eprint {https://arxiv.org/abs/2402.18696}
  {arXiv:2402.18696 [astro-ph.HE]} \BibitemShut {NoStop}%
\bibitem [{\citenamefont {Karkevandi}\ \emph {et~al.}(2022)\citenamefont
  {Karkevandi}, \citenamefont {Shakeri}, \citenamefont {Sagun},\ and\
  \citenamefont {Ivanytskyi}}]{Karkevandi-2021ygv}%
  \BibitemOpen
  \bibfield  {author} {\bibinfo {author} {\bibfnamefont {D.~R.}\ \bibnamefont
  {Karkevandi}}, \bibinfo {author} {\bibfnamefont {S.}~\bibnamefont {Shakeri}},
  \bibinfo {author} {\bibfnamefont {V.}~\bibnamefont {Sagun}},\ and\ \bibinfo
  {author} {\bibfnamefont {O.}~\bibnamefont {Ivanytskyi}},\ }\href
  {https://doi.org/10.1103/PhysRevD.105.023001} {\bibfield  {journal} {\bibinfo
   {journal} {Phys. Rev. D}\ }\textbf {\bibinfo {volume} {105}},\ \bibinfo
  {pages} {023001} (\bibinfo {year} {2022})},\ \Eprint
  {https://arxiv.org/abs/2109.03801} {arXiv:2109.03801 [astro-ph.HE]}
  \BibitemShut {NoStop}%
\bibitem [{\citenamefont {Pisarski}\ and\ \citenamefont
  {Rennecke}(2024)}]{Pisarski:2024esv}%
  \BibitemOpen
  \bibfield  {author} {\bibinfo {author} {\bibfnamefont {R.~D.}\ \bibnamefont
  {Pisarski}}\ and\ \bibinfo {author} {\bibfnamefont {F.}~\bibnamefont
  {Rennecke}},\ }\href {https://doi.org/10.1103/PhysRevLett.132.251903}
  {\bibfield  {journal} {\bibinfo  {journal} {Phys. Rev. Lett.}\ }\textbf
  {\bibinfo {volume} {132}},\ \bibinfo {pages} {251903} (\bibinfo {year}
  {2024})},\ \Eprint {https://arxiv.org/abs/2401.06130} {arXiv:2401.06130
  [hep-ph]} \BibitemShut {NoStop}%
\bibitem [{\citenamefont {Pisarski}\ and\ \citenamefont
  {Rennecke}(2020)}]{Pisarski-2019upw}%
  \BibitemOpen
  \bibfield  {author} {\bibinfo {author} {\bibfnamefont {R.~D.}\ \bibnamefont
  {Pisarski}}\ and\ \bibinfo {author} {\bibfnamefont {F.}~\bibnamefont
  {Rennecke}},\ }\href {https://doi.org/10.1103/PhysRevD.101.114019} {\bibfield
   {journal} {\bibinfo  {journal} {Phys. Rev. D}\ }\textbf {\bibinfo {volume}
  {101}},\ \bibinfo {pages} {114019} (\bibinfo {year} {2020})},\ \Eprint
  {https://arxiv.org/abs/1910.14052} {arXiv:1910.14052 [hep-ph]} \BibitemShut
  {NoStop}%
\bibitem [{\citenamefont {Aghaie}\ \emph {et~al.}(2024)\citenamefont {Aghaie},
  \citenamefont {Armando}, \citenamefont {Conaci}, \citenamefont {Dondarini},
  \citenamefont {Matak}, \citenamefont {Panci}, \citenamefont {Sinska},\ and\
  \citenamefont {Ziegler}}]{Aghaie:2024jkj}%
  \BibitemOpen
  \bibfield  {author} {\bibinfo {author} {\bibfnamefont {M.}~\bibnamefont
  {Aghaie}}, \bibinfo {author} {\bibfnamefont {G.}~\bibnamefont {Armando}},
  \bibinfo {author} {\bibfnamefont {A.}~\bibnamefont {Conaci}}, \bibinfo
  {author} {\bibfnamefont {A.}~\bibnamefont {Dondarini}}, \bibinfo {author}
  {\bibfnamefont {P.}~\bibnamefont {Matak}}, \bibinfo {author} {\bibfnamefont
  {P.}~\bibnamefont {Panci}}, \bibinfo {author} {\bibfnamefont
  {Z.}~\bibnamefont {Sinska}},\ and\ \bibinfo {author} {\bibfnamefont
  {R.}~\bibnamefont {Ziegler}},\ }\href
  {https://doi.org/10.1016/j.physletb.2024.138923} {\bibfield  {journal}
  {\bibinfo  {journal} {Phys. Lett. B}\ }\textbf {\bibinfo {volume} {856}},\
  \bibinfo {pages} {138923} (\bibinfo {year} {2024})},\ \Eprint
  {https://arxiv.org/abs/2404.12199} {arXiv:2404.12199 [hep-ph]} \BibitemShut
  {NoStop}%
\bibitem [{\citenamefont {Lu}\ \emph {et~al.}(2022)\citenamefont {Lu},
  \citenamefont {Xu}, \citenamefont {Wen}, \citenamefont {Peng},\ and\
  \citenamefont {Ruggieri}}]{Lu-2022khf}%
  \BibitemOpen
  \bibfield  {author} {\bibinfo {author} {\bibfnamefont {Z.-Y.}\ \bibnamefont
  {Lu}}, \bibinfo {author} {\bibfnamefont {J.-F.}\ \bibnamefont {Xu}}, \bibinfo
  {author} {\bibfnamefont {X.-J.}\ \bibnamefont {Wen}}, \bibinfo {author}
  {\bibfnamefont {G.-X.}\ \bibnamefont {Peng}},\ and\ \bibinfo {author}
  {\bibfnamefont {M.}~\bibnamefont {Ruggieri}},\ }\href
  {https://doi.org/10.1088/1674-1137/ac5513} {\bibfield  {journal} {\bibinfo
  {journal} {Chin. Phys. C}\ }\textbf {\bibinfo {volume} {46}},\ \bibinfo
  {pages} {064104} (\bibinfo {year} {2022})},\ \Eprint
  {https://arxiv.org/abs/2202.07197} {arXiv:2202.07197 [hep-ph]} \BibitemShut
  {NoStop}%
\end{thebibliography}%

\end{document}